\documentclass[aps,prb,twocolumn,amsmath,amssymb,nofootinbib,superscriptaddress,floatfix]{revtex4-1}
\usepackage{amssymb}
\usepackage{mathrsfs}

\pdfoutput=1

\usepackage{amssymb}
\usepackage{color}
\usepackage{graphicx}
\usepackage{multirow}
\usepackage{ulem}
\usepackage{umoline}
\usepackage{lipsum}

\def\beq{\begin{equation}}
\def\eeq{\end{equation}}

\begin{document}

\title{Universal Entanglement Spectra of Gapped One-dimensional Field Theories}
%
\author{Gil Young Cho}
\affiliation{Department of Physics and Institute for Condensed Matter Theory, University of Illinois, 1110 W. Green St., Urbana, Illinois 61801-3080, USA}
\affiliation{Department of Physics, Korea Advanced Institute of Science and Technology, Daejeon 305-701, Korea}
\author{Andreas W.W. Ludwig} 
\affiliation{Physics Department, University of California, Santa Barbara, CA 93106, USA}
\author{Shinsei Ryu}
\affiliation{Department of Physics and Institute for Condensed Matter Theory, University of Illinois, 1110 W. Green St., Urbana, Illinois 61801-3080, USA}

\date{\today}

\begin{abstract}
We discuss the entanglement spectrum   of  the ground state  of a {\it gapped}  (1+1)-dimensional system  in  a phase  near a quantum phase transition.
In particular, in proximity to a  quantum phase transition described by a conformal field theory (CFT), the system is represented by a gapped Lorentz invariant field theory in the ``scaling limit'' (correlation length $\xi$ much larger than microscopic `lattice' scale `$a$'), and can
be thought of as a  CFT perturbed  by a  relevant perturbation. We show that for such (1+1) {\it gapped} Lorentz invariant
field theories in infinite space, 
 the {\it low-lying  entanglement spectrum} obtained by tracing out, say,  left half-infinite space,  is precisely equal to  the physical spectrum of the 
unperturbed
{\it gapless, i.e. conformal}  field theory defined on a {\it finite interval} of length $L_\xi=$ $\log(\xi/a)$ with
certain boundary conditions. 
In particular, the low-lying entanglement spectrum of the  gapped theory  is the 
finite-size
spectrum of a boundary conformal field theory, and is always
discrete and universal.
 Each relevant perturbation, and thus each gapped phase in proximity to the quantum phase transition,  maps into a particular boundary condition.
A similar property has been known to hold for Baxter's Corner Transfer Matrices in a very special class of fine-tuned, namely integrable off-critical
lattice models, for the entire entanglement spectrum and independent of the scaling limit. In contrast, our result applies to completely general
gapped Lorentz invariant theories in the scaling limit, without the requirement of integrability, for the low-lying entanglement spectrum.
 - While the {\it entanglement spectrum} of the ground state of a gapped theory on  a finite interval of length $2R$ with suitable
boundary conditions, 
bipartitioned  into two equal pieces,
turns out to exhibit 
a
crossover between the finite-size spectra
of the
same
CFT  with
 in general
different boundary conditions as the system size $R$ crosses the correlation length  from the \lq {\it critical regime}\rq \ $R \ll \xi$ to the \lq {\it gapped regime}\rq \ $R \gg \xi$,
the {\it physical spectrum} on a  finite interval of length $R$ with the same boundary conditions, on the other hand,  is known to 
undergo a
dramatic reorganization
during the same crossover from
being discrete to being continuous.
\end{abstract}


\maketitle


\section{Introduction and Summary of Results}
\label{LabelIntroductionAndSummary}



Considerations of Quantum Entanglement have provided a great deal 
of insight into the nature of ground\cite{AmicoFazioOsterlohVedralRMP80-2008-517} 
(and excited\cite{BauerNayakManyBodyLocalization}) states of Hamiltonians 
of complex physical
quantum systems. 
While the entanglement entropy is a very useful diagnostic of a quantum state,  
a vastly larger amount of
 information is contained in the spectrum of the reduced density matrix, i.e. 
in the spectrum of the entanglement Hamiltonian
(Eq.\ (\ref{ReducedDensityMatrix}) below).
For example, a particularly useful case in point is the observation 
that the entanglement Hamiltonian 
of a (2+1)-dimensional integer\cite{Ryu2006} 
as well fractional\cite{LiHaldanePRL2008,ChandranHermannsRegnaultBernevigPRB2011,QiKatsuraLudwigPRL2012,DubailReadRezayiPRB2012,SwingleSenthilEntanglementPRB2012} 
quantum Hall state carries a universal fingerprint of an underlying topological phase.
Indeed, this has recently become an important tool for
identifying the nature of phases of a variety of microscopic gapped Hamiltonians 
by computing the entanglement spectrum numerically.\footnote{See e.g. Ref. \onlinecite{CincioVidalPRL2013,BauerCincioKellerDolfiVidalTrebstLudwig,ZhuGongShengPRB2015,ZhuGongHaldaneShengPRB2015}.}

Here we consider the entanglement spectrum of the ground state of a {\it gapped} (1+1) dimensional  system in a phase
near a quantum phase transition. In particular, we consider phases in proximity to a continuous quantum phase transition
with  dynamical critical exponent $z=1$,  which is generally\footnote{under very mild assumptions\cite{PolchinskiScaleImpliesConformal}} 
described by a conformal field theory (CFT). The system near
such a  transition is thus represented by a gapped Lorentz invariant field theory in the scaling limit (correlation length $\xi$ much
larger than microscopic lattice scale\lq$a$\rq), and can be thought of as a  CFT  perturbed by one or more relevant perturbations.
We consider here primarily the case of a single relevant perturbation, described by  a field $\phi$.

In the present paper we discuss the ground state of such a {\it gapped}
 (1+1) dimensional Lorentz invariant field theory in infinite space.  It is well
 known\cite{KabatStrasslerPhysLettB1994} that the entanglement
Hamiltonian for the ground state of such a gapped theory, obtained by tracing out, say,  
 left half-infinite
 space,  is completely local (being the generator of Lorentz boosts).
 In this paper we will show 
 that in general, the low-lying  {\it entanglement spectrum} of such a    {\it gapped}  theory
is the spectrum of the underlying
unperturbed
{\it gapless}, i.e. conformal theory on a finite interval of length $L_\xi=\ln(\xi/a)$  when $\xi/a \gg 1$, 
with  two boundary conditions:
 a ``free'' boundary condition {``$F$''}  (where the system simply ends)
at the left end of the interval corresponding to the  entanglement cut, and  a ``hard-wall''  boundary condition    which we denote by ``$B_\phi$"
at the other, right end of the interval, which  corresponds to an interface of the CFT
with a strongly gapped phase described by a region of the same theory where the strength of
the relevant perturbation ``$\phi$''  is
by some measure
 large (``infinite''). We emphasize that the latter 
(i.e. right) boundary condition
$B_\phi$  thus depends, as indicated,
on the particular relevant perturbation ``$\phi$'',
and thus on the particular gapped phase in proximity of the transition.
The entanglement Hamiltonian of the gapped theory 
is thus the Hamiltonian  of a boundary conformal field theory (BCFT)\cite{CardyBoundaryFusionVerlindNPB324-1998-581}, with these boundary  conditions. [We note in passing that since the entanglement Hamiltonian
is also known\cite{KabatStrasslerPhysLettB1994} to be equal to the generator of Lorentz boosts of the gapped
relativistic
theory as well as to the Hamiltonian of the same theory
subject to a  uniform acceleration 
(i.e., in Rindler space-time), 
both  the boost operator as well  as the 
Hamiltonian 
in Rindler space-time
also possess this boundary CFT spectrum.]

More explicitly, we show that the entanglement Hamiltonian ${\hat H}_E$ defined through 
the reduced density matrix in 
half-infinite space, region $A=$ 
$\mathbb{R}_+=(0,+\infty)$, 
\begin{eqnarray}
\label{ReducedDensityMatrix}
&& \qquad \qquad \qquad {\hat \rho}_{ A} = {1\over {\cal N}} \exp \{ - 2\pi {\hat H}_E\},  \qquad  \qquad \qquad \qquad    \\  \nonumber
&&{\rm  is \ of \ the \ form}  \qquad  \\  \label{MassiveES}
&&  \quad \qquad  \qquad \ \  
{\hat  H}_E =  {\pi \over L} \left ({\hat  L}_0- {c\over 24}\right ),  \qquad \qquad  \qquad \qquad   \\
\label{DEFLxi}
&&{\rm with}  \qquad  \qquad L = L_\xi \equiv \ln (\xi/a). \qquad \qquad  \qquad \qquad  
\end{eqnarray} 
Here  ${\hat L}_0$ is the chiral (say left-moving)
Virasoro generator
(see {Eq.\ (\ref{BoundarySpectrum})}  below for a more
explicit description)
and $c$ denotes the central charge of the unperturbed theory.
The normalization factor of the reduced density matrix reads
\begin{equation}
\label{normalization}
{\cal N} = {\mathrm{Tr}} \  \exp\{ - 2\pi {\hat H}_E\}= \exp\Big\{ - {c\over 6} L - \gamma + ... \Big\}
\end{equation}
where $\gamma$ is a constant\cite{CardyCalabreseBoundaryEntropy,AffleckLudwigBoundaryEntropy}. 
(The ellipsis indicates terms subleading for large $L$.)
{Equation (\ref{MassiveES})} implies that the  spectrum of eigenvalues $\epsilon$  of the   entanglement Hamiltonian ${\hat H}_E$
is 
of the form
\begin{eqnarray}
\label{BoundarySpectrum}
\epsilon - \epsilon_0 &=& {\pi \over L } \  \{ h + n \},
\end{eqnarray}
where 
$\epsilon_0$ denotes  the smallest eigenvalue. Here $h$ runs over a subset of possible conformal weights\cite{BPZ}
(left-moving scaling dimensions) of the CFT,
which is completely
determined\cite{CardyBoundaryFusionVerlindNPB324-1998-581}  by 
the  pair of boundary conditions ``$F$'' and``$B_\phi$''  at the two ends of the interval of length $L$, 
and $n$ are non-negative integers corresponding to what are known as 
 (conformal)  descendants\cite{BPZ}. 
This is the spectrum of a boundary conformal field theory (e.g., the degeneracies of all levels are known explicitly).

It is only the low-lying entanglement spectrum, 
describing the largest contributions in the Schmidt decomposition 
of the reduced density matrix, 
that is in general universal and described by the spectrum of the BCFT discussed above.
The higher-lying spectrum depends in general on non-universal
details. 
At the end of section \ref{Section-DerivationGeneralFormEntangleSpectrum} 
we provide a rough estimate of the excitation energy
$(\epsilon^*-\epsilon_0)$
at
 which the  conformal spectrum given in (\ref{BoundarySpectrum})
is expected to be no longer applicable,  
which is found to be roughly
$(\epsilon^*-\epsilon_0) \approx$ $2 \pi y$. 
Here $y>0$
 is the renormalization
 group (RG) eigenvalue 
of the relevant perturbation $\phi$, 
a number of order unity. 
Since in view of (\ref{BoundarySpectrum})
the level spacing of the low-lying spectrum is $\pi/L$, 
the number of levels belonging to the low-lying part of the spectrum
increases with $L$. 
In section \ref{LabelSectionNumericalResults}
we present numerical results for  the entanglement spectrum
of  a system of gapped  non-interacting fermions, illustrating our general analytical  results.
We also  note that a reasonably large number of low-lying levels of the entanglement Hamiltonian 
is  within the range of  today's numerical tools even for fully  interacting systems 
as seen, e.g., 
from the numerical entanglement spectra obtained for interacting gapless (conformal) field 
theories
 in Ref.\ [\onlinecite{Laeuchli2013}].\footnote{Recently, numerical results for entanglement spectra of a different set of  
(1+1) dimensional systems, albeit for somewhat  smaller system sizes than in  Ref.\ [\onlinecite{Laeuchli2013}], came to our attention - see Ref. [\onlinecite{KimKatsuraTrivediHan-arXiv1512.08597}].}

A similar property as that derived in the present paper for the low-lying entanglement spectrum of a general gapped
 (1+1) dimensional relativistic field theory
in the vicinity of the CFT, has been known to hold (for many years) for an extremely special  and fine-tuned class of theories, namely for
gapped 
`Yang-Baxter'
 integrable lattice models of 
2D classical
Statistical Mechanics.
Specifically, in these systems Baxter's so-called corner transfer matrix 
\cite{BaxterBook} (CTM)
can be viewed as a lattice analogue of the reduced density matrix in half space, 
{${\hat \rho}_{A}$} from Eq.\ (\ref{ReducedDensityMatrix}), 
when suitably translated into entanglement 
language\footnote{
Specifically, in the scaling limit of the integrable lattice model (which depends on a single parameter that  can be
adjusted so that the system is at criticality)   where the correlation length $\xi$ of the lattice model becomes much
larger than the lattice spacing \lq $a$\rq  \ and where the lattice model turns out to be  represented by a very special,
namely 
integrable relativistic quantum field theory, the CTM ${\hat \rho}_{CTM}$ becomes equal to $[{\hat \rho}_{A}]^{1/4}=$
$(1/{\cal N})^{1/4} \exp\{ -(\pi/2) {\hat H}_E\}$, defined in   Eq. (\ref{ReducedDensityMatrix}). The logarithm of the CTM generalizes the entanglement Hamiltonian (times $\pi/2$) 
to classical 2D Statistical Mechanics systems defined on a lattice, i.e. where Lorentz-invariance is absent.}.
The surprising
observation\cite{CardyLesHouches1988,CardyAdvStudPureMath1989}
was then made for a vast number such integrable lattice systems
(see e.g. Ref.\ [\onlinecite{IntegrableLatticeModelsJapanese}, \onlinecite{SaleurBauer1989}]),  
that the entire spectrum of (minus)  the logarithm of the {CTM}, 
which turns out to play
a role analogous\cite{KabatStrasslerPhysLettB1994} to the entanglement Hamiltonian ${\hat H}_E$ in  Eq. (\ref{ReducedDensityMatrix})
of the field theory,
equals the spectrum of a (gapless) CFT in finite size $L$, 
with the exact replacement $L \to \ln (\xi/a)$, where $a$ 
and $\xi$ are the
 lattice spacing
 and the correlation length, 
respectively,  
of the integrable lattice model. 
Due to the fine-tuning arising
 from 
 integrability
 this turns out to hold 
for all 
eigenvalues of (minus the logarithm of)  the  CTM,
and moreover  holds true for all, even small values of $\xi/a$, 
not only in the scaling limit.
The methods that have been used to demonstrate this fact for these integrable 
systems rely on the very special properties of integrable lattice models, 
such as the Yang-Baxter equation.
Clearly, there is no reason for such a miraculous property to hold without the strong
fine-tuning provided by the 
infinite number of conservation laws
present in these integrable systems.
However, what we show in the present paper is 
that
{\it in the scaling limit}
 the {\it low-lying} entanglement spectrum is {\it generically} 
equal to  that of the  underlying gapless 
theory in finite size, and that this is a property  
completely
independent of the requirement of integrability. 
The identity of these two spectra is thus not a property of the very  
restricted and special class of integrable systems, 
but is a completely general property of the entanglement Hamiltonian of 
gapped (1+1) dimensional
relativistic field theories.

In general, the high-lying excitation spectrum of the entanglement Hamiltonian contains  no robust
information  because it is completely
governed by details of the theory on distance scales  comparable to the microscopic length \lq$a$\rq,
which vary from case to case. On the other hand, an integrable system is known to be very special in this regard, in that even the
short distance properties are completely fixed by the infinite number of conservation laws. One way of expressing this fact
is to think of the integrable theory as a fixed point of the renormalization group (RG), here a  CFT,  
perturbed by an infinite sum of terms that are ever more
irrelevant (in the RG sense), with coefficients that are completely fixed by integrability. This notion has been implemented
in practice in the work of Ref. [\onlinecite{LukyanovTerras-SpinChain-NPB654-2003-323}]. At the end of section \ref{Section-DerivationGeneralFormEntangleSpectrum}, and in particular in Appendix \ref{Appendix-Relationship-with-CMT}, we suggest
that by thinking  this way one may view  the  known results for the CTM of the integrable systems within the  context given in the present paper.

Another related focus of attention in the existing literature on 
the entanglement spectrum of gapped (1+1) dimensional
theories has been the distribution of eigenvalues of the entanglement Hamiltonian
in the regime where the eigenvalues become dense so that the distribution is described by a 
{continuous} curve.  
Ref.\ [\onlinecite{OkunishiHeidaAkutsuPRE59-1999-R6227}] numerically 
observed a universal form
of the distribution of entanglement eigenvalues. 
Later it was argued in Ref.\ [\onlinecite{CalabreseLefevrePRA78-2008-032329}] 
that this distribution has a universal form characterized only by the central charge. 
This was supported by numerical work in that same paper as well as in 
Ref.\ [\onlinecite{PollmannMooreNJPhys12-2010-025006}]. 
While these interesting results are related to the discussion in the present paper,
they do not focus on the resolution of the detailed structure of 
the entanglement spectrum 
on the scale of the 
individual levels, including their degeneracies (which are non-trivial),
discussed  in the present paper.
All this detailed structure on the scale of the level spacing
represents a rich amount of universal information contained in the entanglement spectrum
of the gapped relativistic field theory.

\section{Derivation of the entanglement spectrum of the gapped field theory of  half-space}
\label{Section-DerivationGeneralFormEntangleSpectrum}

We now proceed to provide an explicit derivation of the {\it entanglement Hamiltonian}.



\begin{figure}
\begin{center}
\includegraphics[width=\columnwidth]{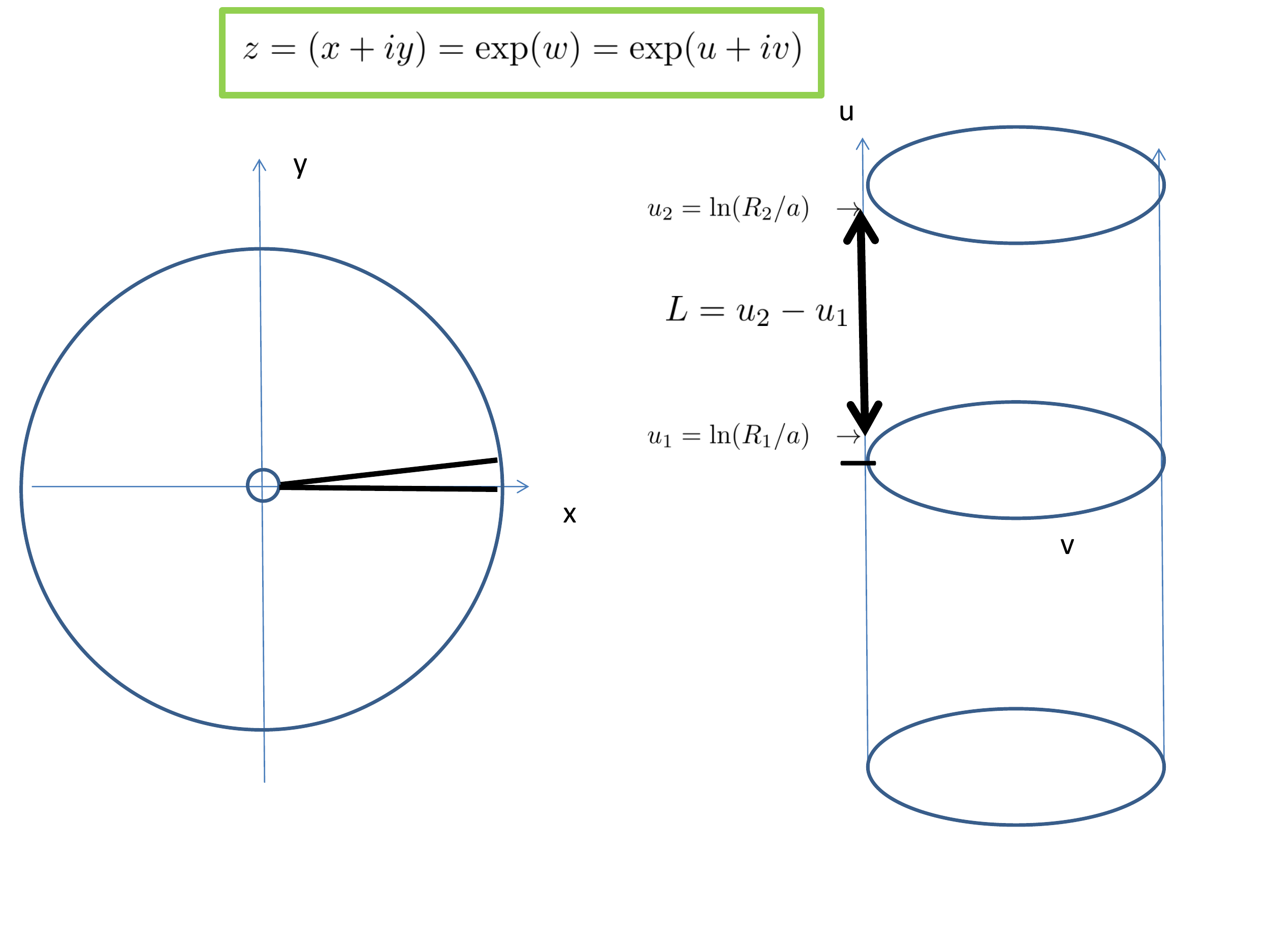}
\caption{
Conformal Map
 \label{Fig0}
}
\end{center}
\end{figure}

We write the spatial coordinate denoted by $x$ and the imaginary (Euclidean) 
time coordinate denoted by $y$ in terms of
$z=x+iy$ and ${\bar z}=x-iy$  
\footnote{
	A characteristic velocity is set to unity for convenience throughout this paper.
}. 
We perform a conformal transformation to a new spatial coordinate $u$ and a new 
imaginary (Euclidean) time coordinate $v$,
via  the conformal transformation 
$z \to w(z)$ where $w=u+iv$ is given by
\footnote{The complex conjugate relationship
holds between ${\bar z}$ and 
${\bar w} = u - i v$.}
\begin{equation}
z=(x+iy) = \exp ( w)= \exp (u+iv),
\label{EuclideanConformalMap}
\end{equation}
mapping the complex $z$-plane into a cylinder - Fig. \ref{Fig0}. 

As it is well known,
there are two equivalent ways of thinking about this transformation: 
(i) as {\it angular quantization} where the angular variable
$v$ is treated as the imaginary (Euclidean) time variable, 
or (ii) as the study of the quantum field theory in Rindler space-time 
which describes the original quantum field theory
subject to a constant acceleration (here  set to unity in suitable units).


Consider the annulus $R_1/a < |z| < R_2/a$,
in the complex $z$-plane (where $a$ is a short distance scale), 
which is mapped (see Fig.\ \ref{Fig0}) under the conformal transformation 
(\ref{EuclideanConformalMap}) 
into a piece  $u_1 < u < u_2$ of a cylinder 
(the coordinate $v$ is periodic with period $2\pi$) of length 
\begin{equation}\label{R1R2}
L = (u_2-u_1) =   \ln (R_2/R_1)
\end{equation}
where
\begin{equation}
\label{Annulus}
R_1/a= \exp(  u_1 ), 
\quad
R_2/a = \exp(   u_2).
\end{equation}

\noindent Now consider, as discussed in the Introduction,  
the imaginary (Euclidean) time  action of a
CFT in the $(x,y)$ coordinate system, perturbed by a 
primary\cite{BPZ}
field
 $\phi(z,{\bar z})$ of conformal weight $(h, {\bar h})$ 
which is relevant in the RG sense,
\begin{equation}\label{ActionRelevantPerturbation}
S_{z, {\bar z}}
=
S_* + g \int d^2 z \ \phi(z, {\bar z}), \quad {\rm where} \ \  {\bar h}=h <1.
\end{equation}
Here $S_*$ denotes the action of the CFT itself.  
$S_{z, {\bar z}}$ in Eq.\ (\ref{ActionRelevantPerturbation})
defines the gapped relativistic field theory in infinite space, 
described by coordinates $(x,y)$
or $(z,{\bar z})$.
In order to obtain its entanglement Hamiltonian in half-space, 
we need to express this
 action in 
$(u,v)$,  or $(w, \bar{w})$ coordinates,
describing angular quantization or, equivalently,  Rindler space-time coordinates.
To this end we use the transformation properties of the primary field $\phi(z, {\bar z})$, which transforms\cite{BPZ}
in the new coordinates to the new field $\Phi(w, {\bar w})$ defined by
\begin{align}
\phi(z, {\bar z})
&=
\Phi(w, {\bar w}) \ \left ( {dz \over dw}\right )^{-h}  \left ( {d {\bar z} \over d {\bar w}} \right )^{-{\bar h}}.
\end{align}
Using the explicit form
(\ref{EuclideanConformalMap})  of the map,
we obtain
$$
\left ( {dz \over dw}\right )  \left ( {d {\bar z} \over d {\bar w}} \right )
=
 \exp (w+ {\bar w})  =
\  \exp (2u  )  
$$
which leads to the following form of the action from (\ref{ActionRelevantPerturbation})  when expressed in the $(w, \bar{w})$ coordinates,
\begin{eqnarray}\nonumber
\label{ActionEuclideanRindlerSpaceTime}
&&S_{w, {\bar w}}
=
S_* + \delta S= \\ \nonumber
&&= 
S_*  
+ \ g \ 
 \int_{u_1}^\infty du \ \int_0^{2\pi} dv
\  \  e^{y  u } \ \  \Phi(w, {\bar w}) = \\
&&=
S_*  
+  
 \int_{u_1}^\infty du \ \int_0^{2\pi} dv
\ e^{y [u-  \ln (\kappa \xi/a)] } \   \Phi(w, {\bar w}),
\end{eqnarray}
where  we made use of the invariance of  $S_*$  under  conformal transformations.
 Here $y=2(1-h)>0$ is  the renormalization group eigenvalue of the relevant coupling  constant $g$,   inducing a finite
(dimensionless) 
correlation length $\xi/a= \kappa^{-1}  \ g^{-1/y}$ (where $\kappa$ is
a non-universal dimensionless constant), which in turn  was  used to write the coupling constant  in the form
$g=$ $\left ( \kappa {\xi \over a} \right )^{-y}=$ $e^{-y \ln (\kappa \xi/a)}$ . We have considered the limit $R_2/a\to \infty$
 (implying 
$u_2\to \infty$, due to Eq. (\ref{Annulus})),
 and
$u_1$ was defined in Eq. (\ref{Annulus}).

Note that the second term $\delta S$ in (\ref{ActionEuclideanRindlerSpaceTime}) 
arises from the presence of the relevant perturbation in (\ref{ActionRelevantPerturbation}) which leads to
the lack of invariance of the total action $S$ under the conformal transformation. 
In the $(w,{\bar w})$ coordinates
the term $\delta S$ describes  a  ``potential'' which grows exponentially with the spatial coordinate $u$ 
and describes an interface between the gapless theory ($g=0$), 
and a gapped theory in which the coupling $g$
is not small, and
 the dimensionless
correlation length $\xi/a$
is not large.
The term $\delta S$
therefore confines the theory to a finite spatial  interval, 
\begin{equation}
\label{FiniteSizeCorrelationLength}
u_1  < u < L, \qquad L = L_\xi=   \ln (\xi/a).
\end{equation}
We thus see from Eqs.\ (\ref{ActionEuclideanRindlerSpaceTime}) and 
(\ref{FiniteSizeCorrelationLength})
that the action $S_{w, \bar w}$ in the $(u,v)$ coordinates of Fig.\ \ref{Fig0} 
describes  
the {\it gapless} theory
but now on a 
space of {\it finite size} $L=$ $L_\xi$, 
with certain
 boundary conditions imposed at the two ends which will be discussed below. 
(Since the imaginary (Euclidean) time coordinate $v$ is
periodic with period $2\pi$, 
this action describes the gapless theory at inverse temperature $\beta=2\pi$.)
Therefore, the Hamiltonian of the theory in the $(u,v)$ coordinates,
which by construction is precisely the entanglement 
Hamiltonian ${\hat H}_E$, 
is
 the Hamiltonian of the underlying  {\it gapless} theory, 
i.e. of the theory where the relevant perturbation is switched off,
$g \equiv 0$, 
but on the {\it finite} interval (\ref{FiniteSizeCorrelationLength}) of 
length $L=$ $L_\xi$. 
The boundary condition on the right end $u=L=$ $L_\xi$ of 
the interval
corresponds, as seen from (\ref{ActionEuclideanRindlerSpaceTime}), 
to an
 interface between the gapless theory (where $g=0$) and 
the fully gapped theory emerging when $g$
is not small.
This interface is sharp
\footnote{The `potential' in
(\ref{ActionEuclideanRindlerSpaceTime}) rises by a factor $e$ when $u$ increases 
by $1/y$ (a number of order unity), 
which is steep as compared to the length $L$ of the interval, when the latter is large.}
when $L$ is large. 
As mentioned in 
section \ref{LabelIntroductionAndSummary},
since the corresponding gapped theory,
appearing when $g$ does not vanish,
clearly depends on the relevant perturbation $\phi$, 
so does the resulting boundary condition, denoted by $B_\phi$. 
On the other hand, the boundary condition at the entanglement cut on the left side $u=u_1$  of the interval is {\it independent} of the  relevant perturbation $\phi$,
and it is  typically\footnote{The ground state typically does not possess  any constraints
between the degrees of freedom immediately on the left ($B$) and the right ($A$) of the entanglement cut. Therefore, upon employing the Schmidt decomposition of the ground state for a bipartition $A\bigcup B$  of  space and  performing the trace over, say,  part $B$
there is no constraint on the leftmost degree of freedom of part $A$; this therefore implies a   ``free'' boundary condition. 
This is also borne out in recent numerical work on the entanglement spectrum of gapless theories, see Ref.\ [\onlinecite{Laeuchli2013}]. - The boundary condition at the entanglement cut can be modified if the ground state contains a specific
constraint on the above-discussed degrees of freedom adjacent to the entanglement cut.}  just a {\it a free} boundary condition (where the system ``simply ends'').

{\it In summary},
we have shown that the low-lying spectrum of the entanglement 
Hamiltonian 
${\hat H}_E$ of the
 gapped relativistic field theory is 
simply the finite size {spectrum} of the corresponding gapless (conformal) 
theory with boundary conditions
{``$F$''} and ``$B_\phi$'' discussed above. 
This is the spectrum of the corresponding {boundary} conformal field theory, 
as displayed in (\ref{MassiveES}), (\ref{DEFLxi}) and (\ref{BoundarySpectrum}). 
The corresponding eigenstates in the $u$-coordinates  are localized within 
the finite range
specified in (\ref{FiniteSizeCorrelationLength}), corresponding in the original $x$-coordinates, upon using (\ref{EuclideanConformalMap}),
as expected
to a finite region around the entanglement cut (at $x=0$)  of spatial extent of the order of the correlation length $\xi$.
- The limitation to the low-lying entanglement spectrum arises from the replacement of the exponentially increasing potential in (\ref{ActionEuclideanRindlerSpaceTime})  by a 
 boundary condition representing a sharp interface. This replacement is certainly
asymptotically valid for  the low-energy, long-wavelength part of   the spectrum 
 when  $L=\ln (\xi/a)$ is large. 
More precisely, one expects this replacement to  stop being valid for eigenstates of the entanglement
Hamiltonian varying on  wavelengths of order $1/y$, 
the scale on which the potential rises exponentially
\footnote{One can check such estimates using the exactly solvable example 
of a quantum mechanical particle in an exponential potential,  
known from Quantum Liouville Theory\cite{GinspargMooreLecturesOn2DGravity-hepth-9304011}.}. 
This replacement is therefore expected to certainly cease to 
be valid for wave vectors $k_n =n (\pi/L)$ with integer $n$ where 
$n \gtrsim  2L y$, which roughly corresponds, using (\ref{BoundarySpectrum}), 
to excitation energies $\epsilon - \epsilon_0 \gtrsim$ $\epsilon_*-\epsilon_0=$ $2\pi y$. 
Since the level spacing is $\pi/L$, the number of energy levels belonging
to the so-defined low-lying spectrum increases with $L$, and can in practice be large in numerical
work (see e.g. Ref. [\onlinecite{Laeuchli2013}], which we mentioned
already
 before; note this reference 
chose to address only  the entanglement spectrum of gapless theories).
A brief comment on how one may view,  within the context of the present paper,  the result observed
for  the logarithm of Baxter's Corner Transfer Matrix in gapped integrable lattice models,
which  is known to reproduce exactly  the entire spectrum of the boundary conformal field theory (i.e. $\epsilon_* \to \infty$
 in the above equation), is provided in
Appendix \ref{Appendix-Relationship-with-CMT}.

\section{Crossover of Entanglement Spectrum of half-space}
\label{LabelSectionAppendixCrossoverEntanglementSpectrum}

 It is illuminating  to compare the {\it finite-size} spectra of the entanglement and the physical Hamiltonian.

\vskip .1cm

\noindent {\it (a): Finite Size Crossover of Entanglement Spectrum.} 
Let us first define  the  finite-size  entanglement spectrum for open boundary conditions  (also often used in numerical work). 
Specifically, we consider the gapped theory on a finite interval $-R < x < +R$,
choosing some, 
for simplicity {\it  identical}   boundary conditions $B_0$ at the two  ends which we assume here to  yield a unique ground state
on the interval. 
When tracing over the  
negative half
 $-R < x <0$ of the interval, we obtain the density matrix whose
spectrum we are interested in.

In the {\it {critical} regime}  where the correlation length is much  larger than the entire interval, $R\ll \xi$, the entanglement spectrum is that of the gapless theory ($g=0$) which is known\cite{PeschelTruong,CardyKITPTalk2015} to be of the form of  Eq. (\ref{BoundarySpectrum}) where
 $L=L_R$ $=\ln (R/a)$:\footnote{This result can be obtained by setting
 $g=0$, $R_2\to R$, $R_1\to a$ and
 $L=L_R=\ln(R/a)$  in  section
\ref{Section-DerivationGeneralFormEntangleSpectrum}.} This is the spectrum of the CFT ($g=0$) on an interval 
of length $L=$ $L_R$ with
typically\footnote{See the corresponding footnote
in the section \ref{Section-DerivationGeneralFormEntangleSpectrum}.}
a  
free boundary condition ``$F$'' on
the left
end (arising from the entanglement cut), and
the same 
 boundary condition 
``$B_0$''  that
 was  imposed in physical space on the right end.
As we increase $R$, the level spacing of the entanglement spectrum
initially decreases as $\pi/\ln (R/a)$, and ultimately saturates at $\pi/\ln(\xi/a)$  
when  we reach the {\it gapped regime}:

In the {\it gapped regime} where the correlation length is much smaller than 
 the
length of the interval,
 $\xi \ll R$, the entanglement 
spectrum is precisely the one studied in section \ref{Section-DerivationGeneralFormEntangleSpectrum}: This is the spectrum of the
same CFT ($g=0$) on an interval on length $L=L_\xi=\ln(\xi/a)$, 
with again  typically a free boundary condition $F$ on the left end (arising from the entanglement cut),
and
the boundary condition $B_\phi$ arising from the relevant perturbation (discussed in section \ref{Section-DerivationGeneralFormEntangleSpectrum}) imposed  on the right end.

We thus see that upon  increasing the size of the interval from $R \ll \xi$ to $\xi \ll R$, the entanglement spectrum
evolves from the spectrum of the CFT on an interval 
 of length 
$L=$ $L_R=\ln(R/a)$ and
 boundary conditions ($F$, $B_0$),   to
the spectrum of the same CFT on an interval
 of length $L=$  $L_\xi=\ln(\xi/a)$ and
 boundary conditions ($F$, $B_\phi$). This describes
the evolution of a boundary renormalization group (RG) flow,
while
 the bulk theory describing the entanglement Hamiltonian always
{remains gapless}.  - A simple example is provided by the transverse field quantum  Ising model
perturbed by a bulk magnetic field described by the operator $\phi=$ $\sigma$ (spin field), and a free 
 Ising-spin
 boundary condition $B_0$.  Upon the above-described crossover of the entanglement spectrum, 
the bulk magnetic field induces a boundary magnetic field at the free spin boundary condition $B_0$ that flows under the RG to the
new fixed-spin boundary condition $B_\phi$. 
- In subsection \ref{LabelSubSectionTopTrivialPhase} of  section \ref{LabelSectionNumericalResults} we discuss  an example
where the  boundary conditions $B_0$ and $B_\phi$ are in fact the same, and only the level spacing changes  upon the crossover. This is confirmed numerically  in
 Fig. \ref{Numerics1}.

\vskip .1cm

\noindent {\it (b): Finite Size Crossover of  Physical Spectrum.}
In contrast to the boundary RG flow of the entanglement spectrum discussed above, the {\it physical spectrum}
is known to evolve completely differently under the analogous crossover. 
In particular consider the  {\it physical spectrum}  of the
gapped theory defined on a finite  interval of length $R$. In order to be able to  make a direct
comparison we choose the same boundary conditions as those for the entanglement spectrum, namely a free boundary condition $F$
on the left end and the boundary condition  $B_0$ on the right end of the interval.

In the {\it critical limit}
 where the correlation length is much  larger than the entire interval, $R\ll \xi$, the physical spectrum
is identical to that  of entanglement 
spectrum, 
Eq. (\ref{BoundarySpectrum}); namely, it is the  spectrum of the CFT
 with the same boundary conditions ($F$, $B_0$), 
upon making the replacement $L _R=\ln(R/a) \to R$.
Upon crossover to the corresponding {\it gapped limit} where the correlation length is much
smaller than the system size, 
$\xi \ll R$ however, 
the physical spectrum of the gapped theory undergoes a dramatic, highly non-trivial re-organization
from the boundary CFT spectrum with a finite level spacing to the continuous spectrum in infinite space describing
{continuous} single- and multi-particle  states of the gapped field theory.  In cases where the relevant
 perturbation of the CFT defining the  gapped theory is integrable, this reorganization of the  physical
  finite size spectrum has been extensively studied in great detail\footnote{Level crossings seen in the integrable cases as a consequence of the
 additional conservation laws will typically turn into avoided crossings in the generic, {non-integrable} settings.}
by means of the so-called `Truncated Space Conformal Field Theory' approach and the Thermodynamic Bethe Ansatz.\footnote{See e.g. Ref.s \onlinecite{TruncatedSpaceCFT,DoreyEtAlJHEP2000}.}

In summary, the finite size  entanglement spectrum
and the finite size physical spectrum 
exhibit entirely different behavior upon crossover from the critical to the gapped regime.

\section{Entanglement spectrum of a finite interval}

\begin{figure}
\begin{center}
\includegraphics[width=\columnwidth]{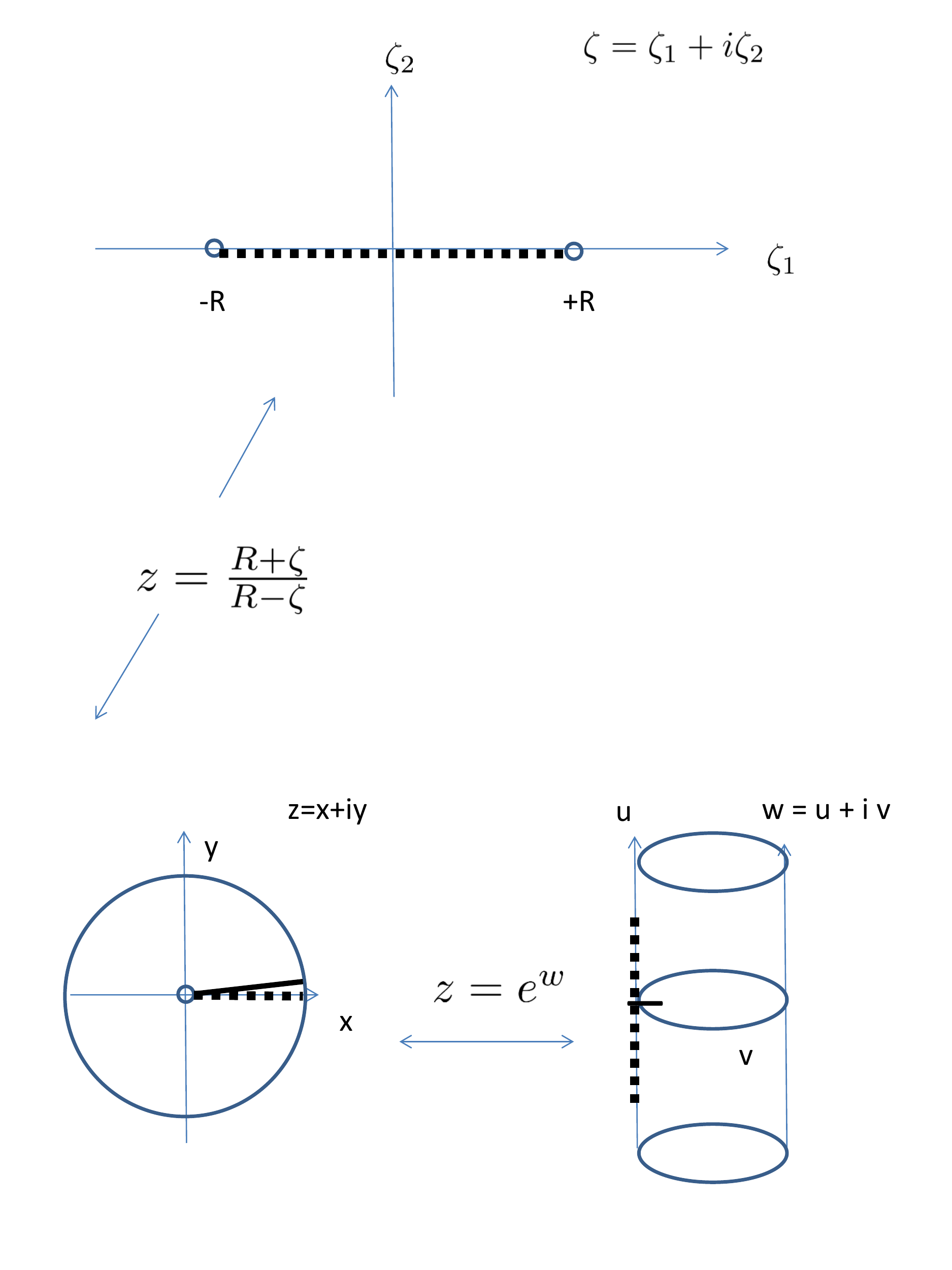}
\caption{
Conformal Map
 \label{Fig1}
}
\end{center}
\end{figure}

For the same gapped field theory, we now discuss the entanglement spectrum of a spatial  interval ${A} =$ $ (-R+a,+R-a)$ where $0 < a\ll R$
(\lq$a$\rq \ is again a short distance scale),
depicted on the real axis with coordinate $\zeta_1$ 
in the top panel of {Fig.\ \ref{Fig1}}. 
We use the conformal map
{$w(\zeta) = \ln \left ({R+\zeta\over R - \zeta }\right )$}, 
with inverse 
$\zeta(w) = R \  \tanh(w/2)$, 
to map from the 
complex $\zeta=$ $(\zeta_1+ i \zeta_2)$-plane 
into a finite cylinder parametrized by 
$w = u + i v$ with  $-u_R \leq u \leq +u_R$ 
where $u_R = \ln (2R/a)$,
as also shown in {Fig.\ \ref{Fig1}}.

As before, consider the imaginary (Euclidean) time action of a CFT 
in the $(\zeta_1,\zeta_2)$ coordinate system, 
perturbed by a
relevant primary\cite{BPZ}
field
 $\phi(\zeta,{\bar \zeta})$
 of conformal weight $(h, {\bar h})$,
$$
S_{\zeta, {\bar \zeta}}
=
S_* + g \int d^2 \zeta \ \phi(\zeta, {\bar \zeta}), \quad {\rm where} \ \  {\bar h}=h <1.
$$
Using
$$
{d\zeta \over dw} = {R\over 2} \ {1\over \cosh(w/2)} ={R\over 1 + \cosh(w)},
$$
we  obtain for the  action in the  $w$-coordinates
\begin{eqnarray}
&& \qquad \qquad  S_{w, {\bar w}}= S_* + \delta S  
\end{eqnarray}
where
\begin{eqnarray}
\nonumber
\label{ActionEuclideanRindlerSpaceTimeZeta}
&&\delta S= \\ \nonumber
&&= g  R^y 
 \int_{-u_R}^{+u_R} du \int_0^{2\pi} dv 
{ 
	 \left[{1\over 1 + \cosh(u)}\right]^y 
 }
\Phi(w, {\bar w})=\\ \nonumber
&&= { ({R\over a})^y\over ({\xi\over a})^y}
\int_{-u_R}^{+u_R} du \ \int_0^{2\pi} dv
{
	\left[{1\over 1 + \cosh(u)}\right]^y
	}
	      	\Phi(w, {\bar w})= \quad \\ \label{FiniteIntervalDeltaS}
&&=
\int_{-u_R}^{+u_R} du \ \int_0^{2\pi} dv
{
	\left[{
e^{(u_R - L_\xi)}\over 1 + \cosh(u)}\right]^y}
  \Phi(w, {\bar w}), \quad 
\end{eqnarray}
and
 $({R\over a})^y=$ $e^{ y  u_R}$.

\noindent In the {\it critical regime} 
{$u_R \ll L_\xi$}, 
where 
$\delta S$ is small  we obtain
the already-known\cite{CardyKITPTalk2015}
entanglement spectrum
of the gapless theory ($g=0$) on the finite interval.\footnote{We see from Fig. 2 that in  the critical regime the entanglement spectrum is the
spectrum of the CFT on a finite interval $-u_R < u < u_R$ with boundary conditions $F$ and $B_\phi$ imposed at the two ends, as in
Ref. \onlinecite{CardyKITPTalk2015}.}

\noindent In the gapped regime $u_R \gg L_\xi$, on the other hand, the ``effective $u$-dependent coupling constant'' in 
the last equation of (\ref{FiniteIntervalDeltaS})  is 
never small
unless $|u|$ is  `close' to $u_R$,
\begin{eqnarray}
\label{SplitIntoTwoIntervals}
(u_R-|u|) \ll L_\xi, \quad ({\rm recall:}   -u_R < u < +u_R),
\end{eqnarray}
in which  case the expression $e^{(u_R - L_\xi)}/[1 + \cosh(u)]$ appearing in Eq. (\ref{ActionEuclideanRindlerSpaceTimeZeta})
tends to  $\to e^{-[L_\xi-(u_R-|u|)]}$ which is small when  $L_\xi$ is large.
The condition in Eq. (\ref{SplitIntoTwoIntervals}) 
 for $\delta S$ to be small
 describes 
two disjoint intervals,
$-u_R < u \ll  -(u_R - L_\xi)$
and
$ (u_R- L_\xi) \ll     u < u_R$; these are  two segments of
length   $L_\xi$ ($\ll u_R$) each
at the right- and the left-
ends  
of the full  interval $-u_R < u < u_R$ in which $u$ is defined.
Therefore, 
the entanglement spectrum of the 
interval ${A}=$ $(-R+a, +R-a)$ in this regime
is  the {\it sum of  two finite size spectra} of the corresponding gapless (conformal)  theory
on a space of 
 length $L_\xi=$ $\ln (\xi/a)$ each. The boundary condition at the ends  $u=\pm u_R$ of each of these two intervals  is (typically) ``free'' $F$,
whereas it is $B_\phi$ at the other ends of the two intervals. Therefore,
each of these two spectra is precisely the spectrum 
discussed in section \ref{Section-DerivationGeneralFormEntangleSpectrum}.


\section{Numerical results}
\label{LabelSectionNumericalResults}


In this section we
present
numerical calculations of 
the (low-lying)  entanglement spectrum of a chain of spinless fermions, 
which
confirm
 the above discussion of Lorentz 
invariant quantum field theories.

Let us consider the Su-Schrieffer-Heeger (SSH) model defined on a one dimensional lattice:
\begin{align}
H=\sum_{i}
	\Psi_{i}^{\dag}h^{0}\Psi_{i}
+\sum_{i}
\left(
	\Psi_{i+1}^{\dag}h^{x}\Psi_{i}+h.c.
\right),
\end{align}
where $i$ labels a site on a one-dimensional chain,
and
$\Psi$ is a two-component fermion 
annihilation operator, which includes two fermions
operators $c_{A}$ and $c_{B}$ defined for each two-site unit cell:
\begin{align}
\Psi_{i}=\left(\begin{array}{c}
c_{A}\\
c_{B}
\end{array}\right)_{i}.
\end{align}
The matrix elements $h^{0}$ and $h^{x}$ are given by: 
\begin{align}
h^{0}=\left(\begin{array}{cc}
\mu_{s} & t+\delta t\\
t+\delta t & -\mu_{s}
\end{array}\right),\quad h^{x}=\left(\begin{array}{cc}
0 & t\\
0 & 0
\end{array}\right).
\end{align}
where $t$ ("hopping"), $\delta t$ ("dimerization"),
and $\mu_s$ ("staggered chemical potential") are real parameters.
In addition, a proper boundary condition must be specified (see below).
For convenience, we will choose 
$
t=1
$
and change $\delta t$ and $\mu_{s}$ 
(which are properly normalized with respect to the unit $t=1$). 
When $\mu_{s}=0$, the SSH
model is particle-hole symmetric, and can be thought of as a member
of symmetry class D.[\onlinecite{AltlandZirnbauer1997},\onlinecite{Schnyder2008}] 
We will mostly set $\mu_{s}=0$ in the following.
Hence, when $\mu_{s}=0$, the SSH model realizes two topologically
distinct gapped phases which are distinguished by a $\mathbb{\mathbb{Z}}_{2}$-valued
topological invariant. 
A topologically trivial phase is realized when
\begin{align}
\delta t>0
\end{align}
while a topologically non-trivial phase is realized when
\begin{align}
\delta t<0.
\end{align}
There is a quantum phase transition separating these phases when 
\begin{align}
\delta t=0.
\end{align}
In the following, we compute the entanglement spectrum of 
these
 phases and at  the critical point between them numerically.

\subsection{Open boundary conditions (OBC)
}

We first consider the SSH model defined on a finite lattice consisting
of $N$ unit cells, and with open boundary condition imposed on both
ends. We will take $N\in 2 \mathbb{Z}$ for convenience.


\subsubsection{ Topologically trivial phase}
\label{LabelSubSectionTopTrivialPhase}

\begin{figure}
\includegraphics[scale=0.9]{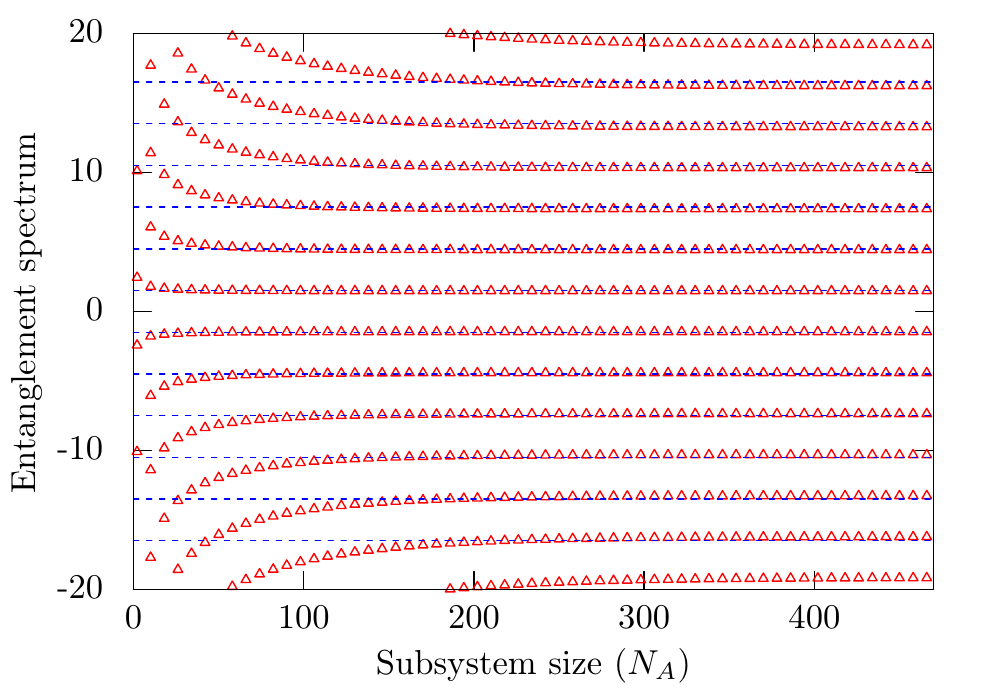}\caption{The 
single-particle
entanglement spectrum of the SSH model with OBC in its topologically trivial
phase ($\delta t=0.01$). The blue dotted lines are guide for eyes.
}
\label{Numerics1}
\end{figure}

Let us start with the topologically trivial phase, $\delta t>0$.
In this phase, the SSH model with open boundary condition has a unique ground state, $|\Psi\rangle$.
We then consider a (pure) density matrix 
$\rho=|\Psi\rangle\langle \Psi|$
made out of the ground state, and 
trace out
the left
 $N/2$ sites to define
the reduced density matrix for the remaining subsystem ("subsystem A"). 
The number of  unit cells in  subsystem A
is denoted by $N_{A}(=N/2)$. 
The computed 
single-particle
entanglement spectrum is presented
in Fig.\ \ref{Numerics1}. 
As we make the system size (and hence the subsystem size)
bigger, 
the spectrum
 approaches 
the prediction
from 
BCFT.
I.e., the levels of the single particle entanglement spectrum
are all equally spaced, and the level spacing does not scale
with  $N_A$.
This spectrum is the spectrum of the free chiral fermion conformal field theory 
with anti-periodic spatial boundary condition,
the so-called Neveu-Schwartz (NS)
spectrum.

\subsubsection{Topologically non-trivial phase}

Next let us consider the non-trivial topological phase, $\delta t<0$. In this
case, the SSH model of finite length
has near double-degenerate ground states when 
open boundary conditions (OBC) are
imposed,
due to near zero-energy single particle modes localized 
near the ends.\footnote{
There is a small splitting, exponential in the system size divided by the correlation length, due to the  interaction (tunneling) between
the zero modes at the two ends of the finite system.}
of the degeneracy, 
we have several options 
to define the density matrix and hence the entanglement spectrum. 
One option would be to take a proper linear combination of the two degenerate ground
states. 
For example, one can make the linear combination such that
the (near) zero-energy single particle eigen state is localized at
a given end. 
One motivation for taking
 such a  linear combination
 is
that
the so-constructed state
may well be compared with the ground
state defined for a semi-infinite system -- the geometry that we considered
in the bulk of the 
paper for the entanglement Hamiltonian.

In practice,
such a ground state
 can be constructed by turning on a
small $\mu_{s}$ (near a boundary, say). 
One should however note that such a
procedure breaks particle-hole symmetry. 
In fact, for any finite system size and finite correlation length, 
if we take a linear combination of near zero-energy modes 
to respect particle-hole symmetry,
they are not localized
at
a given end. 
(Note, however that there is one exception for this: 
the ``zero correlation length limit" 
that we can take in the SSH model). 
Only in the semi-infinite limit, 
(one of) the localized zero energy mode is an exact particle-hole
symmetry eigen state of the Hamiltonian. 

\begin{figure}
\includegraphics[scale=0.9]{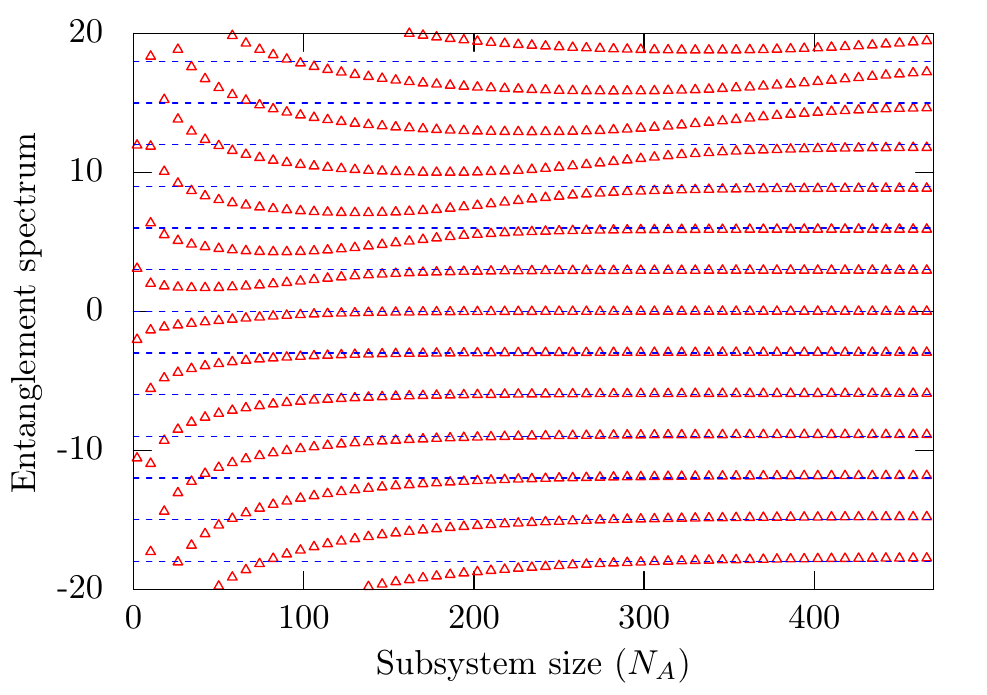}\caption{The entanglement spectrum of the SSH model with OBC in its topologically non-trivial
phase ($\delta t=-0.01$). The blue dotted lines are guide for eyes.}
\label{Numerics2}
\end{figure}

In Fig.\ \ref{Numerics2}, 
the entanglement spectrum with this construction of the ground state
(i.e., the unique ground state selected by 
turning on a finite $\mu_s$)
is shown. 
As we make the subsystem size bigger,
the entanglement spectrum "crosses over".
In particular, 
while for small $N_A$, the single-particle entanglement spectrum
does not have a zero mode,
as we make the system size bigger, 
one level approaches zero from below, 
and asymptotically the single-particle entanglement spectrum has one exact zero mode.  
We call this entanglement spectrum  the Ramond (R)
spectrum 
(the spectrum
 of the free chiral fermion conformal field theory with periodic spatial 
boundary condition).
The R-spectrum, 
whose many-body spectrum displays
 a double degeneracy, 
is what is predicted 
from BCFT (in the \lq gapped regime\rq).

Let now us discuss the ``crossover" in more detail.
In fact, one could discuss two kinds of features separately;
one in terms of the scaling of the level spacing, 
and the other in terms of the structure of the levels, or more precisely
the  presence or absence 
of the zero mode.

From the former perspective, 
if a well-defined crossover region ever appears, 
the spectrum should follow the critical scaling, $\sim H_{L}/\log N_A$,
for $N_A$ much smaller than
the 
correlation length $\xi$
(\lq critical regime\rq).
On the other hand, for $N_A$ much larger than the correlation length, 
the entanglement spectrum should scale as 
$\sim H_{L}/\log\xi$
(\lq gapped regime\rq).
However, it is not entirely
 clear how  to identify
such a crossover
region in Fig.\ \ref{Numerics2}.

From the perspective of the structure of the spectrum,
the entanglement spectrum for small $N_A$ looks like the NS spectrum
(i.e., there is no zero mode in the single particle entanglement spectrum),
which crosses over to the R-spectrum.
(However, this "NS-like" spectrum should be 
distinguished 
from 
the NS-spectrum that appears in the topologically trivial case,
because of the different scaling of the level spacing.)
It is here crucial to recall that, by our construction of the unique ground state, 
the entanglement spectrum breaks particle-hole symmetry. 
In fact, what would look 
like
a crossover in numerics is possible because
of the particle-hole symmetry breaking.
This should be compared with what we expect for the ideal, 
semi-infinite limit. 
In the semi-infinite limit, 
the ground state is unique and respects particle-hole symmetry,
and so is the reduced density matrix.
If so, the entanglement spectrum should be particle-hole symmetric. 
As a corrollary, the spectrum
cannot cross over
 from the NS {spectrum} to R spectrum in the presence of particle-hole symmetry. 
In short, the crossover from the NS {spectrum} to R spectrum in
numerics is due to particle-hole symmetry breaking,
which we can think of as a finite size artifact.
On the other hand, as we make the system size
bigger,  particle-hole symmetry breaking eventually goes away,
and hence, in this ideal limit, the entanglement spectrum will be particle-hole
symmetric. 
There should
in fact
 be no crossover from
an  actual
 NS to
the
 R spectrum. 
\footnote{
We should remark that the crossover (from small to large size $N_A$) 
observed in the  numerical (single-particle) entanglement spectrum displayed in Fig. \ref{Numerics2} is not quite covered by the cases discussed 
analytically in 
section \ref{LabelSectionAppendixCrossoverEntanglementSpectrum}. This is because in 
section \ref{LabelSectionAppendixCrossoverEntanglementSpectrum} for simplicity
the technical assumption was made that the finite
interval, to be bipartitioned into two pieces, had {\it identical} boundary conditions (called $B_0$ in that section)
imposed at the two ends. (This permitted the simple  analytical prediction  of the crossover
of the entanglement spectrum  detailed in section \ref{LabelSectionAppendixCrossoverEntanglementSpectrum}.)
. For the case where {\it different} boundary conditions are imposed at the two ends,
the corresponding entanglement spectrum
has not yet been studied using analytical means. The numerical study presented in Fig. \ref{Numerics2} of such a case
may help the development of a future analytical description of this type of crossover of the entanglement spectrum.
}

\subsection{
Periodic boundary conditions (PBC)}

\begin{figure}
\includegraphics[scale=0.9]{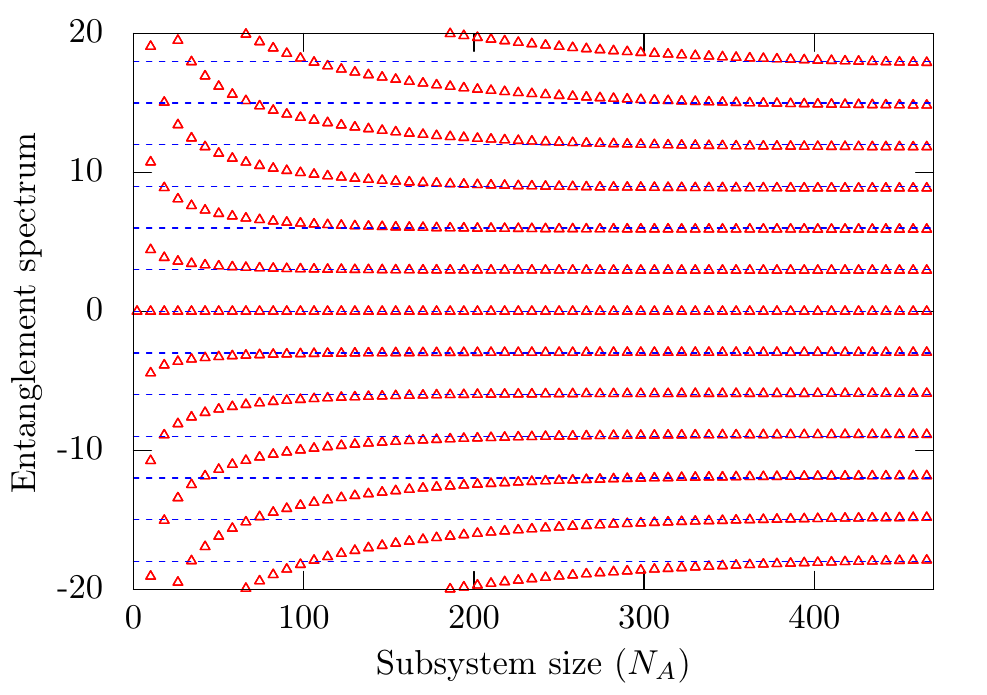}
\caption{
	\label{ES PBC}
	The single-particle  entanglement spectrum of the SSH model with PBC 
	in its topologically non-trivial
phase ($\delta t=-0.01$). The blue dotted lines are guide for eyes.
Each level is doubly degenerate.}
\end{figure}

Let us now discuss the entanglement spectrum for the case of 
periodic boundary
conditions.
With periodic boundary 
conditions,
the ground state is always unique as far as there is a spectral gap,
irrespective of the sign of $\delta t$. I.e., even in the topological
case,
one does not have to choose between ground states. 
The numerically computed entanglement spectrum is shown in Fig.\ \ref{ES PBC}. 
As compared to the case of open boundary 
conditions, 
there is no crossover from the NS to the  R spectrum. 
(As mentioned above, such a  crossover is not to be expected in
the ideal semi-infinite limit.) 
One important feature for the case of periodic boundary
conditions,
which we expect from our discussion in the 
preceding section, 
is that,
due to the presence of two entangling boundaries, 
 the entanglement spectrum consists of two idential copies of
a
 BCFT spectrum. 
I.e., each level in the single-particle entanglement spectrum is doubly degenerate. 
The double-degeneracy is indeed confirmed in numerics. 
In this special case of equal bipartition, $N_B=N-N_A=N_A$, 
this double-degeneracy turns out to be exact. 
On the other hand, if, instead of taking $N_{A}=N/2$, 
we choose $1\le N_{A}\le N/2$,
the double degeneracy is lifted. 
This can be understood as an ``tunneling'' (coupling) between the two 
copies of 
the
BCFT
spectrum.

\subsection{Critical scaling}

\begin{figure}
	\includegraphics[scale=0.9]{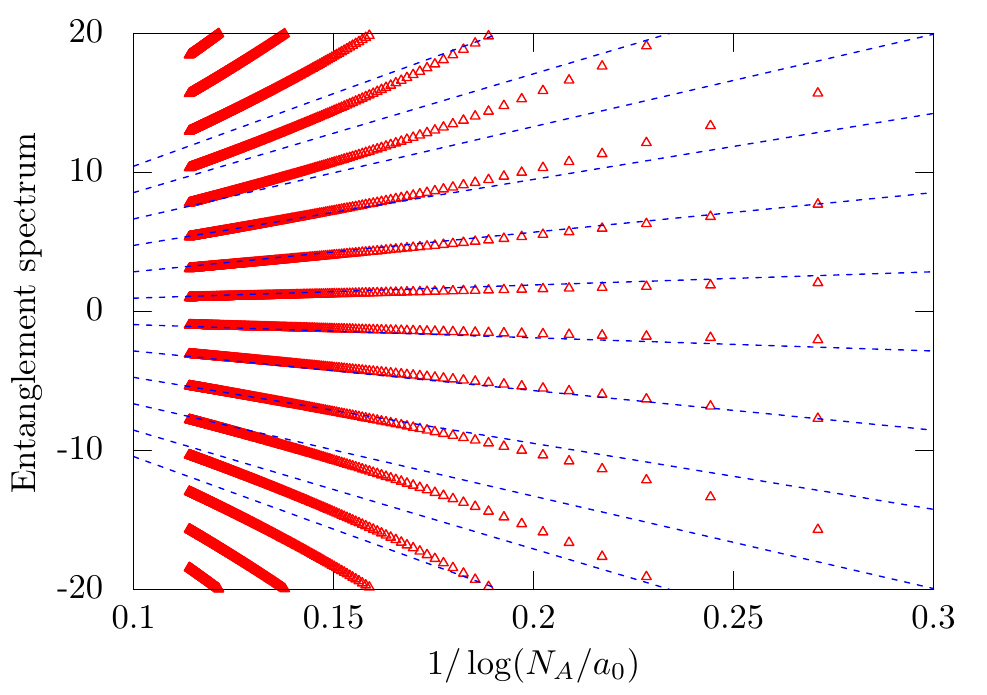}
	\caption{
		\label{ES critical OBC}
		The single-particle  entanglement spectrum of the SSH model with periodic boundary conditions at its critical point ($\delta t=0$).
The blue dotted lines are guide for eyes. To compare with the expected
result, we chose the
\lq scale\rq \ 
 $a_{0}$
to be
 $a_{0}=0.1.$ }
\label{LabelFigCritical}
\end{figure}

Finally, as an aside, let us take a look at the 
entanglement spectrum at the critical point $\delta t=0$. 
From the field theory considerations, we expect that
the spectrum scales as $\sim1/\log N_{A}$ instead of $\sim1/\log\xi.$
In Fig.\ \ref{ES critical OBC},
we fit the entanglement spectrum at the critical point,
which roughly 
follows
 what we expect;
one observes
critical scaling $H_L/\log N_A$ when $N_A$ is large enough. 

In gapless systems (critical points) in general, 
since there is no spectral gap, there are many (near) degenerate states.
In our numerics, we let our
 computers 
choose
 the ground state. 
This procedure in principle may be tricky;
for example, the behavior of the entanglement spectrum may depend severely
on the choice of $N_{A}$ and $N$. (There may be an even-odd like effect. 
In fact, when we choose periodic boundary 
conditions,
there is a noticeable even-odd like effect,
however, there is no such effect for open boundary 
conditions.)

\section{Conclusions}

In conclusion, we have considered in this paper  (1+1) dimensional gapped
relativistic field
 theories
 in the scaling limit which can be viewed as
describing gapped phases 
in  the vicinity of a quantum phase transition described by
a CFT. We have shown that the low-lying  entanglement spectrum of such a field
theory is the spectrum of the underlying CFT on a finite interval of size $L_\xi=\ln(\xi/a)$ with a free boundary
condition $F$ and a boundary $B_\phi$ determined by the relevant perturbation of the CFT
defining the gapped theory. We have also calculated the entanglement spectrum of the gapped field
theory on a finite interval. This result provides, at the same time, the structure of the entanglement
spectrum of the theory with periodic boundary conditions, i.e. on a circle of circumference $2R$, bipartitioned
into two half-circles of length $R$ each: In the limit where the size $R$ is larger than the correlation length $\xi$,
in analogy with the case of an interval,
the entanglement spectrum is the sum of two spectra arising from the two ends of the semicircle.
We would like to mention that interesting features arise for entanglement spectra of (1+1)
dimensional Symmetry Protected Topological (SPT) phases, and these will be discussed in a companion
paper by the authors which will appear very shortly.

\begin{acknowledgments} 

{We thank John Cardy for}
an interesting discussion.
We are grateful to the KITP Program {\it  Entanglement in Strongly-Correlated Quantum Matter}
(Apr 6 - Jul 2, 2015), where some part of the work was performed.
This work is supported by 
the NSF under Grants 
No.\ NSF PHY11-25915,
No.\ DMR-1064319 (GYC), 
and 
No.\ DMR-1309667 (A.W.W.L.),
the Brain Korea 21 PLUS Project of Korea Government (GYC),
and by the  Alfred P. Sloan foundation (SR).


\end{acknowledgments}

\appendix


\section{Comments on the relationship with the Corner Transfer Matrix Spectum of Yang-Baxter integrable 2D classical Statistical Mechanics models}
\label{Appendix-Relationship-with-CMT}

In this Appendix we briefly suggest a way in which the
observation made (many years ago) in the literature about the spectrum of the logarithm of the CTM  of
integrable lattice models  could be viewed within the context of the notions used in the present paper about the low-lying  entanglement
spectrum of gapped relativistic (1+1) dimensional  field theories. This will also  provide some  intuition about how the highly special constraints of
integrability manage to generate the exact BCFT spectrum at arbitrarily high excitation \lq energies\rq  \ of the entanglement Hamiltonian,
even far off the scaling limit, i.e.  for values of the correlation length $\xi$ down to distances of the lattice scale \lq$a$\rq.
First, recall that all the  many  off-critical  integrable lattice  models in which the behavior in question
of the CTM was observed, are in fact one-parameter families of integrable models, which have the property that 
the system is critical ($\xi/a=\infty$)  for one special value $\lambda_*$ of the parameter $\lambda$, 
where they represent  lattice realizations of a certain set of  CFTs. Moreover, the deviation $\delta \lambda =$ $(\lambda-\lambda_*)$
of the parameter from this special value is a relevant perturbation, and couples to a particular relevant field $\phi$ of the CFT. (This
particular field  is also special in the context of the CFT, in that even in the CFT perturbed by it an infinite subset of the conservation laws of the CFT
survive. Only a few very special  relevant perturbations of the CFT have this property.)
Second, recall that (i)   one can represent the critical lattice theory at $\lambda=\lambda_*$ as the CFT perturbed by an infinite number of
irrelevant perturbations. Some of these irrelevant perturbations include powers of the energy-momentum tensor of the CFT
which lead to non-linear contributions to the energy-momentum relationship, thus representing the  breaking of Lorentz-invariance
present in  the lattice theory.
Moreover, (ii),  the off-critical lattice model at $\delta \lambda \not = 0$ may be represented by perturbing the
so-represented critical lattice theory at $\lambda=\lambda_*$ by yet another  infinite set of perturbations of the CFT, 
the most relevant of which is the field $\phi$ discussed above. It is the constraints arising from the integrability of the lattice model
that  fix exactly the infinite number of expansion coefficients.
A practical implementation of this general principle of  representing an integrable  lattice theory in terms of such
perturbations of a CFT, can be found e.g. in Ref. [\onlinecite{LukyanovTerras-SpinChain-NPB654-2003-323}].
- Now it is clear that the presence of all these perturbations (in principle infinite in number)  can be treated  in precisely
the same way the single relevant perturbation $\phi$ is treated in section \ref{Section-DerivationGeneralFormEntangleSpectrum} of the present
paper:
In that section it is shown that a single relevant perturbation $\phi$, when added to the CFT,  leads to a ``domain wall potential''
in the coordinate \lq$u$\rq \  parametrizing the coordinate space on which the entanglement Hamiltonian ${\hat H}_E$ acts (`angular quantization'
or `Rindler space-time'). All the additional perturbations that have to be added to the CFT to represent the lattice theory
exactly will simply modify the {\it shape} of that ``domain wall potential'' in some way.  Since the spectrum of the logarithm of the CTM
of the gapped integrable models is known to be exactly the entire  spectrum of the boundary CFT with the boundary conditions
mentioned  in the present paper, it ought to be  the case that the 
{\it exact} ``domain wall potential'' being generated in this way
in the  integrable theory, {\it exactly} describes a conformal
boundary condition, all the way up to arbitrarily high (entanglement) energies. It may be possible to study explicitly, in the spirit
of  Ref. [\onlinecite{LukyanovTerras-SpinChain-NPB654-2003-323}] mentioned above,
approximations to the entire   entanglement spectrum  of the integrable system
in terms of that of a CFT with a finite, but increasing number of perturbations.






\vskip .1cm

%
%

\bibliography{ref4new}

\begin{thebibliography}{52}%
\makeatletter
\providecommand \@ifxundefined [1]{%
 \@ifx{#1\undefined}
}%
\providecommand \@ifnum [1]{%
 \ifnum #1\expandafter \@firstoftwo
 \else \expandafter \@secondoftwo
 \fi
}%
\providecommand \@ifx [1]{%
 \ifx #1\expandafter \@firstoftwo
 \else \expandafter \@secondoftwo
 \fi
}%
\providecommand \natexlab [1]{#1}%
\providecommand \enquote  [1]{``#1''}%
\providecommand \bibnamefont  [1]{#1}%
\providecommand \bibfnamefont [1]{#1}%
\providecommand \citenamefont [1]{#1}%
\providecommand \href@noop [0]{\@secondoftwo}%
\providecommand \href [0]{\begingroup \@sanitize@url \@href}%
\providecommand \@href[1]{\@@startlink{#1}\@@href}%
\providecommand \@@href[1]{\endgroup#1\@@endlink}%
\providecommand \@sanitize@url [0]{\catcode `\\12\catcode `\$12\catcode
  `\&12\catcode `\#12\catcode `\^12\catcode `\_12\catcode `\%12\relax}%
\providecommand \@@startlink[1]{}%
\providecommand \@@endlink[0]{}%
\providecommand \url  [0]{\begingroup\@sanitize@url \@url }%
\providecommand \@url [1]{\endgroup\@href {#1}{\urlprefix }}%
\providecommand \urlprefix  [0]{URL }%
\providecommand \Eprint [0]{\href }%
\providecommand \doibase [0]{http://dx.doi.org/}%
\providecommand \selectlanguage [0]{\@gobble}%
\providecommand \bibinfo  [0]{\@secondoftwo}%
\providecommand \bibfield  [0]{\@secondoftwo}%
\providecommand \translation [1]{[#1]}%
\providecommand \BibitemOpen [0]{}%
\providecommand \bibitemStop [0]{}%
\providecommand \bibitemNoStop [0]{.\EOS\space}%
\providecommand \EOS [0]{\spacefactor3000\relax}%
\providecommand \BibitemShut  [1]{\csname bibitem#1\endcsname}%
\let\auto@bib@innerbib\@empty
\bibitem [{\citenamefont {Amico}\ \emph {et~al.}(2008)\citenamefont {Amico},
  \citenamefont {Fazio}, \citenamefont {Osterloh},\ and\ \citenamefont
  {Vedral}}]{AmicoFazioOsterlohVedralRMP80-2008-517}%
  \BibitemOpen
  \bibfield  {author} {\bibinfo {author} {\bibfnamefont {L.}~\bibnamefont
  {Amico}}, \bibinfo {author} {\bibfnamefont {R.}~\bibnamefont {Fazio}},
  \bibinfo {author} {\bibfnamefont {A.}~\bibnamefont {Osterloh}}, \ and\
  \bibinfo {author} {\bibfnamefont {V.}~\bibnamefont {Vedral}},\ }\href@noop {}
  {\bibfield  {journal} {\bibinfo  {journal} {Rev. Mod. Phys.}\ }\textbf
  {\bibinfo {volume} {80}},\ \bibinfo {pages} {517} (\bibinfo {year}
  {2008})}\BibitemShut {NoStop}%
\bibitem [{\citenamefont {Bauer}\ and\ \citenamefont
  {Nayak}(2014)}]{BauerNayakManyBodyLocalization}%
  \BibitemOpen
  \bibfield  {author} {\bibinfo {author} {\bibfnamefont {B.}~\bibnamefont
  {Bauer}}\ and\ \bibinfo {author} {\bibfnamefont {C.}~\bibnamefont {Nayak}},\
  }\href@noop {} {\bibfield  {journal} {\bibinfo  {journal} {J. Stat. Mech.}\
  }\textbf {\bibinfo {volume} {2013}},\ \bibinfo {pages} {P09005} (\bibinfo
  {year} {2014})}\BibitemShut {NoStop}%
\bibitem [{\citenamefont {{Ryu}}\ and\ \citenamefont
  {{Hatsugai}}(2006)}]{Ryu2006}%
  \BibitemOpen
  \bibfield  {author} {\bibinfo {author} {\bibfnamefont {S.}~\bibnamefont
  {{Ryu}}}\ and\ \bibinfo {author} {\bibfnamefont {Y.}~\bibnamefont
  {{Hatsugai}}},\ }\href {\doibase 10.1103/PhysRevB.73.245115} {\bibfield
  {journal} {\bibinfo  {journal} {\prb}\ }\textbf {\bibinfo {volume} {73}},\
  \bibinfo {eid} {245115} (\bibinfo {year} {2006})},\ \Eprint
  {http://arxiv.org/abs/cond-mat/0601237} {cond-mat/0601237} \BibitemShut
  {NoStop}%
\bibitem [{\citenamefont {Li}\ and\ \citenamefont
  {Haldane}(2008)}]{LiHaldanePRL2008}%
  \BibitemOpen
  \bibfield  {author} {\bibinfo {author} {\bibfnamefont {H.}~\bibnamefont
  {Li}}\ and\ \bibinfo {author} {\bibfnamefont {F.}~\bibnamefont {Haldane}},\
  }\href@noop {} {\bibfield  {journal} {\bibinfo  {journal} {Phys. Rev. Lett.}\
  }\textbf {\bibinfo {volume} {101}},\ \bibinfo {pages} {010504} (\bibinfo
  {year} {2008})}\BibitemShut {NoStop}%
\bibitem [{\citenamefont {Chandran}\ \emph {et~al.}(2011)\citenamefont
  {Chandran}, \citenamefont {Hermanns}, \citenamefont {Regnault},\ and\
  \citenamefont {Bernevig}}]{ChandranHermannsRegnaultBernevigPRB2011}%
  \BibitemOpen
  \bibfield  {author} {\bibinfo {author} {\bibfnamefont {A.}~\bibnamefont
  {Chandran}}, \bibinfo {author} {\bibfnamefont {M.}~\bibnamefont {Hermanns}},
  \bibinfo {author} {\bibfnamefont {N.}~\bibnamefont {Regnault}}, \ and\
  \bibinfo {author} {\bibfnamefont {B.}~\bibnamefont {Bernevig}},\ }\href@noop
  {} {\bibfield  {journal} {\bibinfo  {journal} {Physical Review B}\ }\textbf
  {\bibinfo {volume} {84}},\ \bibinfo {pages} {205136} (\bibinfo {year}
  {2011})}\BibitemShut {NoStop}%
\bibitem [{\citenamefont {Qi}\ \emph {et~al.}(2012)\citenamefont {Qi},
  \citenamefont {Katsura},\ and\ \citenamefont
  {Ludwig}}]{QiKatsuraLudwigPRL2012}%
  \BibitemOpen
  \bibfield  {author} {\bibinfo {author} {\bibfnamefont {X.-L.}\ \bibnamefont
  {Qi}}, \bibinfo {author} {\bibfnamefont {H.}~\bibnamefont {Katsura}}, \ and\
  \bibinfo {author} {\bibfnamefont {A.}~\bibnamefont {Ludwig}},\ }\href@noop {}
  {\bibfield  {journal} {\bibinfo  {journal} {Physical Review Letters}\
  }\textbf {\bibinfo {volume} {108}},\ \bibinfo {pages} {196402} (\bibinfo
  {year} {2012})}\BibitemShut {NoStop}%
\bibitem [{\citenamefont {Dubail}\ \emph {et~al.}(2012)\citenamefont {Dubail},
  \citenamefont {Read},\ and\ \citenamefont
  {Rezayi}}]{DubailReadRezayiPRB2012}%
  \BibitemOpen
  \bibfield  {author} {\bibinfo {author} {\bibfnamefont {J.}~\bibnamefont
  {Dubail}}, \bibinfo {author} {\bibfnamefont {N.}~\bibnamefont {Read}}, \ and\
  \bibinfo {author} {\bibfnamefont {E.}~\bibnamefont {Rezayi}},\ }\href@noop {}
  {\bibfield  {journal} {\bibinfo  {journal} {Physical Review B}\ }\textbf
  {\bibinfo {volume} {86}},\ \bibinfo {pages} {245310} (\bibinfo {year}
  {2012})}\BibitemShut {NoStop}%
\bibitem [{\citenamefont {Swingle}\ and\ \citenamefont
  {Senthil}(2012)}]{SwingleSenthilEntanglementPRB2012}%
  \BibitemOpen
  \bibfield  {author} {\bibinfo {author} {\bibfnamefont {B.}~\bibnamefont
  {Swingle}}\ and\ \bibinfo {author} {\bibfnamefont {T.}~\bibnamefont
  {Senthil}},\ }\href@noop {} {\bibfield  {journal} {\bibinfo  {journal}
  {Physical Review B}\ }\textbf {\bibinfo {volume} {86}},\ \bibinfo {pages}
  {045117} (\bibinfo {year} {2012})}\BibitemShut {NoStop}%
\bibitem [{Note1()}]{Note1}%
  \BibitemOpen
  \bibinfo {note} {See e.g. Ref. \protect \rev@citealpnum
  {CincioVidalPRL2013,BauerCincioKellerDolfiVidalTrebstLudwig,ZhuGongShengPRB2015,ZhuGongHaldaneShengPRB2015}.}\BibitemShut
  {Stop}%
\bibitem [{Note2()}]{Note2}%
  \BibitemOpen
  \bibinfo {note} {Under very mild assumptions\cite
  {PolchinskiScaleImpliesConformal}}\BibitemShut {NoStop}%
\bibitem [{\citenamefont {Kabat}\ and\ \citenamefont
  {Strassler}(1994)}]{KabatStrasslerPhysLettB1994}%
  \BibitemOpen
  \bibfield  {author} {\bibinfo {author} {\bibfnamefont {D.}~\bibnamefont
  {Kabat}}\ and\ \bibinfo {author} {\bibfnamefont {M.}~\bibnamefont
  {Strassler}},\ }\href@noop {} {\bibfield  {journal} {\bibinfo  {journal}
  {Physics Letters B}\ }\textbf {\bibinfo {volume} {329}},\ \bibinfo {pages}
  {46} (\bibinfo {year} {1994})}\BibitemShut {NoStop}%
\bibitem [{\citenamefont
  {Cardy}(1989{\natexlab{a}})}]{CardyBoundaryFusionVerlindNPB324-1998-581}%
  \BibitemOpen
  \bibfield  {author} {\bibinfo {author} {\bibfnamefont {J.}~\bibnamefont
  {Cardy}},\ }\href@noop {} {\bibfield  {journal} {\bibinfo  {journal} {Nucl.
  Phys. B}\ }\textbf {\bibinfo {volume} {324}},\ \bibinfo {pages} {581}
  (\bibinfo {year} {1989}{\natexlab{a}})}\BibitemShut {NoStop}%
\bibitem [{\citenamefont {Calabrese}\ and\ \citenamefont
  {Cardy}(2002)}]{CardyCalabreseBoundaryEntropy}%
  \BibitemOpen
  \bibfield  {author} {\bibinfo {author} {\bibfnamefont {P.}~\bibnamefont
  {Calabrese}}\ and\ \bibinfo {author} {\bibfnamefont {J.}~\bibnamefont
  {Cardy}},\ }\href@noop {} {\bibfield  {journal} {\bibinfo  {journal} {J.
  Stat. Mech.}\ }\textbf {\bibinfo {volume} {2002}},\ \bibinfo {pages} {P06002}
  (\bibinfo {year} {2002})}\BibitemShut {NoStop}%
\bibitem [{\citenamefont {Affleck}\ and\ \citenamefont
  {Ludwig}(1991)}]{AffleckLudwigBoundaryEntropy}%
  \BibitemOpen
  \bibfield  {author} {\bibinfo {author} {\bibfnamefont {I.}~\bibnamefont
  {Affleck}}\ and\ \bibinfo {author} {\bibfnamefont {A.}~\bibnamefont
  {Ludwig}},\ }\href@noop {} {\bibfield  {journal} {\bibinfo  {journal} {Phys.
  Rev. Lett.}\ }\textbf {\bibinfo {volume} {67}},\ \bibinfo {pages} {161}
  (\bibinfo {year} {1991})}\BibitemShut {NoStop}%
\bibitem [{\citenamefont {Belavin}\ \emph {et~al.}(1984)\citenamefont
  {Belavin}, \citenamefont {Polyakov},\ and\ \citenamefont
  {Zamolodchikov}}]{BPZ}%
  \BibitemOpen
  \bibfield  {author} {\bibinfo {author} {\bibfnamefont {A.}~\bibnamefont
  {Belavin}}, \bibinfo {author} {\bibfnamefont {A.~M.}\ \bibnamefont
  {Polyakov}}, \ and\ \bibinfo {author} {\bibfnamefont {A.~B.}\ \bibnamefont
  {Zamolodchikov}},\ }\href@noop {} {\bibfield  {journal} {\bibinfo  {journal}
  {Nucl. Phys. B}\ }\textbf {\bibinfo {volume} {241}},\ \bibinfo {pages} {333}
  (\bibinfo {year} {1984})}\BibitemShut {NoStop}%
\bibitem [{\citenamefont {Laeuchli}(2013)}]{Laeuchli2013}%
  \BibitemOpen
  \bibfield  {author} {\bibinfo {author} {\bibfnamefont {A.}~\bibnamefont
  {Laeuchli}},\ }\href@noop {} {\bibfield  {journal} {\bibinfo  {journal}
  {arXiv-1303.0741}\ } (\bibinfo {year} {2013})}\BibitemShut {NoStop}%
\bibitem [{Note3()}]{Note3}%
  \BibitemOpen
  \bibinfo {note} {Recently, numerical results for entanglement spectra of a
  different set of (1+1) dimensional systems, albeit for somewhat smaller
  system sizes than in Ref.\ [\protect \rev@citealpnum {Laeuchli2013}], came to
  our attention - see Ref. [\protect \rev@citealpnum
  {KimKatsuraTrivediHan-arXiv1512.08597}].}\BibitemShut {Stop}%
\bibitem [{\citenamefont {Baxter}(1982)}]{BaxterBook}%
  \BibitemOpen
  \bibfield  {author} {\bibinfo {author} {\bibfnamefont {R.~J.}\ \bibnamefont
  {Baxter}},\ }\href@noop {} {}{Exactly Solved Models in Statistical
  Mechancis}\ (\bibinfo  {publisher} {Academic},\ \bibinfo {year}
  {1982})\BibitemShut {NoStop}%
\bibitem [{Note4()}]{Note4}%
  \BibitemOpen
  \bibinfo {note} {Specifically, in the scaling limit of the integrable lattice
  model (which depends on a single parameter that can be adjusted so that the
  system is at criticality) where the correlation length $\xi $ of the lattice
  model becomes much larger than the lattice spacing `$a$'\ and where the
  lattice model turns out to be represented by a very special, namely
  integrable relativistic quantum field theory, the CTM ${\protect \mathaccentV
  {hat}05E\rho }_{CTM}$ becomes equal to $[{\protect \mathaccentV {hat}05E\rho
  }_{A}]^{1/4}=$ $(1/{\protect \cal N})^{1/4} \protect \qopname \relax
  o{exp}\protect \{ -(\pi /2) {\protect \mathaccentV {hat}05EH}_E\protect \}$,
  defined in Eq. (\ref {ReducedDensityMatrix}). The logarithm of the CTM
  generalizes the entanglement Hamiltonian (times $\pi /2$) to classical 2D
  Statistical Mechanics systems defined on a lattice, i.e. where
  Lorentz-invariance is absent.}\BibitemShut {Stop}%
\bibitem [{\citenamefont {Cardy}(1990)}]{CardyLesHouches1988}%
  \BibitemOpen
  \bibfield  {author} {\bibinfo {author} {\bibfnamefont {J.}~\bibnamefont
  {Cardy}},\ }\href@noop {} {\bibfield  {journal} {\bibinfo  {journal} {in: Les
  Houches Summer School 1988 {\it Fields, Strings and Critical Phenomena}, eds.
  E.Br\'ezin and J.Zinn-Justin;}\ }\textbf {\bibinfo {volume} {XLIX}},\
  \bibinfo {pages} {169} (\bibinfo {year} {1990})}\BibitemShut {NoStop}%
\bibitem [{\citenamefont
  {Cardy}(1989{\natexlab{b}})}]{CardyAdvStudPureMath1989}%
  \BibitemOpen
  \bibfield  {author} {\bibinfo {author} {\bibfnamefont {J.~L.}\ \bibnamefont
  {Cardy}},\ }\href@noop {} {\bibfield  {journal} {\bibinfo  {journal}
  {Advanced Studies in Pure Mathematics}\ }\textbf {\bibinfo {volume} {19}},\
  \bibinfo {pages} {127} (\bibinfo {year} {1989}{\natexlab{b}})}\BibitemShut
  {NoStop}%
\bibitem [{\citenamefont {Date}\ \emph {et~al.}(1987)\citenamefont {Date},
  \citenamefont {Jimbo}, \citenamefont {Miwa},\ and\ \citenamefont
  {M.}}]{IntegrableLatticeModelsJapanese}%
  \BibitemOpen
  \bibfield  {author} {\bibinfo {author} {\bibfnamefont {E.}~\bibnamefont
  {Date}}, \bibinfo {author} {\bibfnamefont {M.}~\bibnamefont {Jimbo}},
  \bibinfo {author} {\bibfnamefont {T.}~\bibnamefont {Miwa}}, \ and\ \bibinfo
  {author} {\bibfnamefont {O.}~\bibnamefont {M.}},\ }\href@noop {} {\bibfield
  {journal} {\bibinfo  {journal} {Physical Review B}\ }\textbf {\bibinfo
  {volume} {35}},\ \bibinfo {pages} {2101} (\bibinfo {year}
  {1987})}\BibitemShut {NoStop}%
\bibitem [{\citenamefont {Saleur}\ and\ \citenamefont
  {Bauer}(1989)}]{SaleurBauer1989}%
  \BibitemOpen
  \bibfield  {author} {\bibinfo {author} {\bibfnamefont {H.}~\bibnamefont
  {Saleur}}\ and\ \bibinfo {author} {\bibfnamefont {M.}~\bibnamefont {Bauer}},\
  }\href@noop {} {\bibfield  {journal} {\bibinfo  {journal} {Nucl. Phys.}\
  }\textbf {\bibinfo {volume} {B324}},\ \bibinfo {pages} {581} (\bibinfo {year}
  {1989})}\BibitemShut {NoStop}%
\bibitem [{\citenamefont {Lukyanov}\ and\ \citenamefont
  {Terras}(2003)}]{LukyanovTerras-SpinChain-NPB654-2003-323}%
  \BibitemOpen
  \bibfield  {author} {\bibinfo {author} {\bibfnamefont {S.}~\bibnamefont
  {Lukyanov}}\ and\ \bibinfo {author} {\bibfnamefont {V.}~\bibnamefont
  {Terras}},\ }\href@noop {} {\bibfield  {journal} {\bibinfo  {journal} {Nucl.
  Phys.}\ }\textbf {\bibinfo {volume} {B654}},\ \bibinfo {pages} {323}
  (\bibinfo {year} {2003})}\BibitemShut {NoStop}%
\bibitem [{\citenamefont {Okunishi}\ \emph {et~al.}(1999)\citenamefont
  {Okunishi}, \citenamefont {Hieida},\ and\ \citenamefont
  {Akutsu}}]{OkunishiHeidaAkutsuPRE59-1999-R6227}%
  \BibitemOpen
  \bibfield  {author} {\bibinfo {author} {\bibfnamefont {K.}~\bibnamefont
  {Okunishi}}, \bibinfo {author} {\bibfnamefont {Y.}~\bibnamefont {Hieida}}, \
  and\ \bibinfo {author} {\bibfnamefont {Y.}~\bibnamefont {Akutsu}},\
  }\href@noop {} {\bibfield  {journal} {\bibinfo  {journal} {Phys. Rev. E}\
  }\textbf {\bibinfo {volume} {59}},\ \bibinfo {pages} {R6227} (\bibinfo {year}
  {1999})}\BibitemShut {NoStop}%
\bibitem [{\citenamefont {Calabrese}\ and\ \citenamefont
  {Lefevre}(2008)}]{CalabreseLefevrePRA78-2008-032329}%
  \BibitemOpen
  \bibfield  {author} {\bibinfo {author} {\bibfnamefont {P.}~\bibnamefont
  {Calabrese}}\ and\ \bibinfo {author} {\bibfnamefont {A.}~\bibnamefont
  {Lefevre}},\ }\href@noop {} {\bibfield  {journal} {\bibinfo  {journal} {Phys.
  Rev. A}\ }\textbf {\bibinfo {volume} {78}},\ \bibinfo {pages} {032329}
  (\bibinfo {year} {2008})}\BibitemShut {NoStop}%
\bibitem [{\citenamefont {Pollmann}\ and\ \citenamefont
  {Moore}(2010)}]{PollmannMooreNJPhys12-2010-025006}%
  \BibitemOpen
  \bibfield  {author} {\bibinfo {author} {\bibfnamefont {F.}~\bibnamefont
  {Pollmann}}\ and\ \bibinfo {author} {\bibfnamefont {J.~E.}\ \bibnamefont
  {Moore}},\ }\href@noop {} {\bibfield  {journal} {\bibinfo  {journal} {N.J.
  Phys}\ }\textbf {\bibinfo {volume} {12}},\ \bibinfo {pages} {025006}
  (\bibinfo {year} {2010})}\BibitemShut {NoStop}%
\bibitem [{Note5()}]{Note5}%
  \BibitemOpen
  \bibinfo {note} {A characteristic velocity is set to unity for convenience
  throughout this paper.}\BibitemShut {Stop}%
\bibitem [{Note6()}]{Note6}%
  \BibitemOpen
  \bibinfo {note} {The complex conjugate relationship holds between ${\protect
  \mathaccentV {bar}016z}$ and ${\protect \mathaccentV {bar}016w} = u - i
  v$.}\BibitemShut {Stop}%
\bibitem [{Note7()}]{Note7}%
  \BibitemOpen
  \bibinfo {note} {The `potential' in (\ref {ActionEuclideanRindlerSpaceTime})
  rises by a factor $e$ when $u$ increases by $1/y$ (a number of order unity),
  which is steep as compared to the length $L$ of the interval, when the latter
  is large.}\BibitemShut {Stop}%
\bibitem [{Note8()}]{Note8}%
  \BibitemOpen
  \bibinfo {note} {The ground state typically does not possess any constraints
  between the degrees of freedom immediately on the left ($B$) and the right
  ($A$) of the entanglement cut. Therefore, upon employing the Schmidt
  decomposition of the ground state for a bipartition $A\DOTSB \bigcup@
  \slimits@ B$ of space and performing the trace over, say, part $B$ there is
  no constraint on the leftmost degree of freedom of part $A$; this therefore
  implies a ``free'' boundary condition. This is also borne out in recent
  numerical work on the entanglement spectrum of gapless theories, see Ref.\
  [\protect \rev@citealpnum {Laeuchli2013}]. - The boundary condition at the
  entanglement cut can be modified if the ground state contains a specific
  constraint on the above-discussed degrees of freedom adjacent to the
  entanglement cut.}\BibitemShut {Stop}%
\bibitem [{Note9()}]{Note9}%
  \BibitemOpen
  \bibinfo {note} {One can check such estimates using the exactly solvable
  example of a quantum mechanical particle in an exponential potential, known
  from Quantum Liouville Theory\cite
  {GinspargMooreLecturesOn2DGravity-hepth-9304011}.}\BibitemShut {Stop}%
\bibitem [{\citenamefont {Peschel}\ and\ \citenamefont
  {Truong}(1987)}]{PeschelTruong}%
  \BibitemOpen
  \bibfield  {author} {\bibinfo {author} {\bibfnamefont {I.}~\bibnamefont
  {Peschel}}\ and\ \bibinfo {author} {\bibfnamefont {T.~T.}\ \bibnamefont
  {Truong}},\ }\href@noop {} {\bibfield  {journal} {\bibinfo  {journal} {Z.
  Physik}\ }\textbf {\bibinfo {volume} {B 69}},\ \bibinfo {pages} {391}
  (\bibinfo {year} {1987})}\BibitemShut {NoStop}%
\bibitem [{\citenamefont {Cardy}()}]{CardyKITPTalk2015}%
  \BibitemOpen
  \bibfield  {author} {\bibinfo {author} {\bibfnamefont {J.}~\bibnamefont
  {Cardy}},\ }\href@noop {} {}{Talk, KITP, June 2015: On the (unmetaphorical)
  entanglement gap in CFTs (and related topics) }\BibitemShut {NoStop}%
\bibitem [{Note10()}]{Note10}%
  \BibitemOpen
  \bibinfo {note} {This result can be obtained by setting $g=0$, $R_2\to R$,
  $R_1\to a$ and $L=L_R=\protect \qopname \relax o{ln}(R/a)$ in section \ref
  {Section-DerivationGeneralFormEntangleSpectrum}.}\BibitemShut {Stop}%
\bibitem [{Note11()}]{Note11}%
  \BibitemOpen
  \bibinfo {note} {See the corresponding footnote in the section \ref
  {Section-DerivationGeneralFormEntangleSpectrum}.}\BibitemShut {Stop}%
\bibitem [{Note12()}]{Note12}%
  \BibitemOpen
  \bibinfo {note} {Level crossings seen in the integrable cases as a
  consequence of the additional conservation laws will typically turn into
  avoided crossings in the generic, {non-integrable} settings.}\BibitemShut
  {Stop}%
\bibitem [{Note13()}]{Note13}%
  \BibitemOpen
  \bibinfo {note} {See e.g. Ref.s \protect \rev@citealpnum
  {TruncatedSpaceCFT,DoreyEtAlJHEP2000}.}\BibitemShut {Stop}%
\bibitem [{Note14()}]{Note14}%
  \BibitemOpen
  \bibinfo {note} {We see from Fig. 2 that in the critical regime the
  entanglement spectrum is the spectrum of the CFT on a finite interval $-u_R <
  u < u_R$ with boundary conditions $F$ and $B_\phi $ imposed at the two ends,
  as in Ref. \protect \rev@citealpnum {CardyKITPTalk2015}.}\BibitemShut {Stop}%
\bibitem [{\citenamefont {{Altland}}\ and\ \citenamefont
  {{Zirnbauer}}(1997)}]{AltlandZirnbauer1997}%
  \BibitemOpen
  \bibfield  {author} {\bibinfo {author} {\bibfnamefont {A.}~\bibnamefont
  {{Altland}}}\ and\ \bibinfo {author} {\bibfnamefont {M.~R.}\ \bibnamefont
  {{Zirnbauer}}},\ }\href@noop {} {\bibfield  {journal} {\bibinfo  {journal}
  {Physical Review B}\ }\textbf {\bibinfo {volume} {55}},\ \bibinfo {eid}
  {4986} (\bibinfo {year} {1997})}\BibitemShut {NoStop}%
\bibitem [{\citenamefont {Schnyder}\ \emph {et~al.}(2008)\citenamefont
  {Schnyder}, \citenamefont {Ryu}, \citenamefont {Furusaki},\ and\
  \citenamefont {Ludwig}}]{Schnyder2008}%
  \BibitemOpen
  \bibfield  {author} {\bibinfo {author} {\bibfnamefont {A.~P.}\ \bibnamefont
  {Schnyder}}, \bibinfo {author} {\bibfnamefont {S.}~\bibnamefont {Ryu}},
  \bibinfo {author} {\bibfnamefont {A.}~\bibnamefont {Furusaki}}, \ and\
  \bibinfo {author} {\bibfnamefont {A.~W.~W.}\ \bibnamefont {Ludwig}},\ }\href
  {http://link.aps.org/doi/10.1103/PhysRevB.78.195125} {\bibfield  {journal}
  {\bibinfo  {journal} {Phys. Rev. B}\ }\textbf {\bibinfo {volume} {78}},\
  \bibinfo {pages} {195125} (\bibinfo {year} {2008})}\BibitemShut {NoStop}%
\bibitem [{Note15()}]{Note15}%
  \BibitemOpen
  \bibinfo {note} {There is a small splitting, exponential in the system size
  divided by the correlation length, due to the interaction (tunneling) between
  the zero modes at the two ends of the finite system.}\BibitemShut {Stop}%
\bibitem [{Note16()}]{Note16}%
  \BibitemOpen
  \bibinfo {note} {We should remark that the crossover (from small to large
  size $N_A$) observed in the numerical (single-particle) entanglement spectrum
  displayed in Fig. \ref {Numerics2} is not quite covered by the cases
  discussed analytically in section \ref
  {LabelSectionAppendixCrossoverEntanglementSpectrum}. This is because in
  section \ref {LabelSectionAppendixCrossoverEntanglementSpectrum} for
  simplicity the technical assumption was made that the finite interval, to be
  bipartitioned into two pieces, had {\protect \it identical} boundary
  conditions (called $B_0$ in that section) imposed at the two ends. (This
  permitted the simple analytical prediction of the crossover of the
  entanglement spectrum detailed in section \ref
  {LabelSectionAppendixCrossoverEntanglementSpectrum}.) . For the case where
  {\protect \it different} boundary conditions are imposed at the two ends, the
  corresponding entanglement spectrum has not yet been studied using analytical
  means. The numerical study presented in Fig. \ref {Numerics2} of such a case
  may help the development of a future analytical description of this type of
  crossover of the entanglement spectrum.}\BibitemShut {Stop}%
\bibitem [{\citenamefont {Cincio}\ and\ \citenamefont
  {Vidal}(2013)}]{CincioVidalPRL2013}%
  \BibitemOpen
  \bibfield  {author} {\bibinfo {author} {\bibfnamefont {L.}~\bibnamefont
  {Cincio}}\ and\ \bibinfo {author} {\bibfnamefont {G.}~\bibnamefont {Vidal}},\
  }\href@noop {} {\bibfield  {journal} {\bibinfo  {journal} {Physical Review
  Letters}\ }\textbf {\bibinfo {volume} {110}},\ \bibinfo {pages} {067208}
  (\bibinfo {year} {2013})}\BibitemShut {NoStop}%
\bibitem [{\citenamefont {Bauer}\ \emph {et~al.}(2014)\citenamefont {Bauer},
  \citenamefont {Cincio}, \citenamefont {Keller}, \citenamefont {Dolfi},
  \citenamefont {Vidal}, \citenamefont {Trebst},\ and\ \citenamefont
  {Ludwig}}]{BauerCincioKellerDolfiVidalTrebstLudwig}%
  \BibitemOpen
  \bibfield  {author} {\bibinfo {author} {\bibfnamefont {B.}~\bibnamefont
  {Bauer}}, \bibinfo {author} {\bibfnamefont {L.}~\bibnamefont {Cincio}},
  \bibinfo {author} {\bibfnamefont {B.}~\bibnamefont {Keller}}, \bibinfo
  {author} {\bibfnamefont {M.}~\bibnamefont {Dolfi}}, \bibinfo {author}
  {\bibfnamefont {G.}~\bibnamefont {Vidal}}, \bibinfo {author} {\bibfnamefont
  {S.}~\bibnamefont {Trebst}}, \ and\ \bibinfo {author} {\bibfnamefont
  {A.}~\bibnamefont {Ludwig}},\ }\href@noop {} {\bibfield  {journal} {\bibinfo
  {journal} {Nature Communications}\ }\textbf {\bibinfo {volume} {5}},\
  \bibinfo {pages} {5137} (\bibinfo {year} {2014})}\BibitemShut {NoStop}%
\bibitem [{\citenamefont {Zhu}\ \emph {et~al.}(2015{\natexlab{a}})\citenamefont
  {Zhu}, \citenamefont {Gong},\ and\ \citenamefont
  {Sheng}}]{ZhuGongShengPRB2015}%
  \BibitemOpen
  \bibfield  {author} {\bibinfo {author} {\bibfnamefont {W.}~\bibnamefont
  {Zhu}}, \bibinfo {author} {\bibfnamefont {S.~S.}\ \bibnamefont {Gong}}, \
  and\ \bibinfo {author} {\bibfnamefont {D.~N.}\ \bibnamefont {Sheng}},\
  }\href@noop {} {\bibfield  {journal} {\bibinfo  {journal} {Physical Review
  B}\ }\textbf {\bibinfo {volume} {92}},\ \bibinfo {pages} {014424} (\bibinfo
  {year} {2015}{\natexlab{a}})}\BibitemShut {NoStop}%
\bibitem [{\citenamefont {Zhu}\ \emph {et~al.}(2015{\natexlab{b}})\citenamefont
  {Zhu}, \citenamefont {Gong}, \citenamefont {Haldane},\ and\ \citenamefont
  {Sheng}}]{ZhuGongHaldaneShengPRB2015}%
  \BibitemOpen
  \bibfield  {author} {\bibinfo {author} {\bibfnamefont {W.}~\bibnamefont
  {Zhu}}, \bibinfo {author} {\bibfnamefont {S.~S.}\ \bibnamefont {Gong}},
  \bibinfo {author} {\bibfnamefont {F.~D.~M.}\ \bibnamefont {Haldane}}, \ and\
  \bibinfo {author} {\bibfnamefont {D.~N.}\ \bibnamefont {Sheng}},\ }\href@noop
  {} {\bibfield  {journal} {\bibinfo  {journal} {Physical Review B}\ }\textbf
  {\bibinfo {volume} {92}},\ \bibinfo {pages} {165106} (\bibinfo {year}
  {2015}{\natexlab{b}})}\BibitemShut {NoStop}%
\bibitem [{\citenamefont {Polchinski}(1988)}]{PolchinskiScaleImpliesConformal}%
  \BibitemOpen
  \bibfield  {author} {\bibinfo {author} {\bibfnamefont {J.}~\bibnamefont
  {Polchinski}},\ }\href@noop {} {\bibfield  {journal} {\bibinfo  {journal}
  {Nucl. Phys. B}\ }\textbf {\bibinfo {volume} {303}},\ \bibinfo {pages} {226}
  (\bibinfo {year} {1988})}\BibitemShut {NoStop}%
\bibitem [{\citenamefont {Kim}\ \emph {et~al.}(2015)\citenamefont {Kim},
  \citenamefont {Katsura}, \citenamefont {Trivedi},\ and\ \citenamefont
  {Han}}]{KimKatsuraTrivediHan-arXiv1512.08597}%
  \BibitemOpen
  \bibfield  {author} {\bibinfo {author} {\bibfnamefont {P.}~\bibnamefont
  {Kim}}, \bibinfo {author} {\bibfnamefont {H.}~\bibnamefont {Katsura}},
  \bibinfo {author} {\bibfnamefont {N.}~\bibnamefont {Trivedi}}, \ and\
  \bibinfo {author} {\bibfnamefont {J.}~\bibnamefont {Han}},\ }\href@noop {}
  {\bibfield  {journal} {\bibinfo  {journal} {arXiv-1512.08597}\ } (\bibinfo
  {year} {2015})}\BibitemShut {NoStop}%
\bibitem [{\citenamefont {Ginsparg}\ and\ \citenamefont
  {Moore}(1993)}]{GinspargMooreLecturesOn2DGravity-hepth-9304011}%
  \BibitemOpen
  \bibfield  {author} {\bibinfo {author} {\bibfnamefont {P.}~\bibnamefont
  {Ginsparg}}\ and\ \bibinfo {author} {\bibfnamefont {G.}~\bibnamefont
  {Moore}},\ }\href@noop {} {\bibfield  {journal} {\bibinfo  {journal} {ArXiv
  e-prints}\ } (\bibinfo {year} {1993})},\ \Eprint
  {http://arxiv.org/abs/9304911} {arXiv:9304911} \BibitemShut {NoStop}%
\bibitem [{\citenamefont {Yurov}\ and\ \citenamefont
  {Zamolodchikov}(1990)}]{TruncatedSpaceCFT}%
  \BibitemOpen
  \bibfield  {author} {\bibinfo {author} {\bibfnamefont {V.~P.}\ \bibnamefont
  {Yurov}}\ and\ \bibinfo {author} {\bibfnamefont {A.~B.}\ \bibnamefont
  {Zamolodchikov}},\ }\href@noop {} {\bibfield  {journal} {\bibinfo  {journal}
  {Int. J. Mod. Phys.}\ }\textbf {\bibinfo {volume} {A5}},\ \bibinfo {pages}
  {3221} (\bibinfo {year} {1990})}\BibitemShut {NoStop}%
\bibitem [{\citenamefont {Dorey}\ \emph {et~al.}()\citenamefont {Dorey},
  \citenamefont {Pillin}, \citenamefont {Pocklington}, \citenamefont {Runkel},
  \citenamefont {Tateo},\ and\ \citenamefont {Watts}}]{DoreyEtAlJHEP2000}%
  \BibitemOpen
  \bibfield  {author} {\bibinfo {author} {\bibfnamefont {P.}~\bibnamefont
  {Dorey}}, \bibinfo {author} {\bibfnamefont {M.}~\bibnamefont {Pillin}},
  \bibinfo {author} {\bibfnamefont {A.}~\bibnamefont {Pocklington}}, \bibinfo
  {author} {\bibfnamefont {I.}~\bibnamefont {Runkel}}, \bibinfo {author}
  {\bibfnamefont {R.}~\bibnamefont {Tateo}}, \ and\ \bibinfo {author}
  {\bibfnamefont {G.}~\bibnamefont {Watts}},\ }\href@noop {} {\ }\Eprint
  {http://arxiv.org/abs/0010278} {arXiv:0010278} \BibitemShut {NoStop}%
\end{thebibliography}%

\end{document}